\documentclass[traditabstract]{aa}
\usepackage{natbib}
\bibpunct{(}{)}{;}{a}{}{,}
\usepackage{revsymb}    
\usepackage{amsmath}    
\usepackage[usenames]{color}    
\newcommand{\derivp} [2] {\frac {\partial #1 } {\partial #2} }
\newcommand{\deriv} [2] {\frac {\textrm{d} #1 } {\textrm{d} #2} }
\newcommand{\eq}[1] {Eq.\,(\ref{#1})}

\usepackage{txfonts}

\begin{document}

\title{A non-local mixing-length theory able to compute core overshooting}

\author{M. Gabriel\inst{1} \and K. Belkacem\inst{2}}

\institute{
Institut d'Astrophysique et de G\'eophysique, Universit\'e de Li\`ege, All\'ee du 6 Ao\^ut 17-B 4000 Li\`ege, Belgium 
\and
LESIA, Observatoire de Paris, CNRS, PSL Research University, Universit\'e Pierre et Marie Curie, Universit\'e Denis Diderot, 92195 Meudon, France
}

   \offprints{M. Gabriel}
   \mail{fb101462@skynet.be}
   \date{\today}

  \authorrunning{M. Gabriel \& K. Belkacem}

   \abstract{Turbulent convection is certainly one of the most important and thorny issues in stellar physics. Our deficient knowledge of this crucial physical process introduces a fairly large uncertainty concerning the internal structure and evolution of stars. A striking example is overshoot at the edge of convective cores. Indeed, nearly all stellar evolutionary codes treat the overshooting zones in a very approximative way that considers both its extent and the profile of the temperature gradient as free parameters. There are only a few sophisticated theories of stellar convection such as Reynolds stress approaches, but they also require the adjustment of a non-negligible number of free parameters. We present here a theory, based on the plume theory as well as on the mean-field equations, but without relying on the usual Taylor’s closure hypothesis. It leads us to a set of eight differential equations plus a few algebraic ones. Our theory is essentially a non-mixing length theory. It enables us to compute the temperature gradient in a shrinking convective core and its overshooting zone. The case of an expanding convective core is also discussed, though more briefly.  
   Numerical simulations have quickly improved during recent years and enabling us to foresee that they will probably soon provide a model of convection adapted to the computation of 1D stellar models. 
   }


   \maketitle

\section{Introduction}
\label{intro} 

Convective overshoot is a long-standing and thorny problem in stellar physics. \cite{Shaviv1973} were the first to try to compute the overshooting  above a convective core. They computed the slowdown of a rising convective bubble above the boundary, as given by the Schwarzschild criterion, assuming that the temperature gradient is slightly sub-adiabatic. However, they did not discuss the motion of the downward  material, which remained unexplained. Their work was followed by many others, which are critically discussed in \cite{Renzini1987}. Due to the large uncertainties of the theories available at that time, the overshooting distance was quickly considered as a free parameter, taken proportional to the pressure scale height at the classical surface of the core (i.e. as given by the classical Schwarzschild criterion). In addition, there is still no agreement on the value of the temperature gradient to be adopted in the overshooting layers. Some codes assume an adiabatic gradient while others prefer the radiative one, so that the extent obtained through fitting of evolutionary tracks using observations depends on this choice. 

From the late seventies, the group led by Xiong \citep[see][and references therein]{Xiong2001} developed a theory of convection based on a Reynolds stress approach and were able to incorporate it into an evolution code \citep[see also the more recent work by][]{Kupka2002}. This type of approach is quite powerful and widely used in geophysics but it provides results that are very sensitive to the adopted closure model (and associated free parameters) as well as to the domain of applicability of the closure model. Consequently, only a handful of evolutions have been computed using it. A very complete review giving the present state of the problem and including a clear historic of the modelling of convection is given by \cite{Kupka2017}. 
Recently, based on 3D hydrodynamical simulations of convection in massive stars \citep[e.g.][]{Meakin2007a,Arnett2009,Arnett2011,Viallet2013}, \cite{Arnett2015} proposed a model based on a Reynolds-averaged Navier–Stokes equations (RANS) analysis to propose a set of equations, which they call the 321D algorithm, to include in 1D evolution codes. This promising approach presents the advantage of solving the problem and of dealing with convective boundary layers without making an assumption concerning the flow geometry. 

On the other hand, \cite{Schmitt1984} introduced the theory of turbulent plumes in astrophysics. It was applied to stellar envelopes by \cite{RieutordZahn1995} and adapted to convective cores by \cite{LoSchatzman1997}. In these works, the spherically-averaged equations are ignored with the exception of the continuity equation, which is used to modify the advection term given by Taylor’s hypothesis. This assumption, also known as the turbulent entrainment hypothesis, states that the radial inflow of matter is proportional to the central vertical velocity inside a plume \citep[see][for details]{Turner1986}. 
This approach has several consequences, the main one being that the coefficient of proportionality in Taylor's hypothesis is to be specified. It is commonly considered as a universal value but has only been constrained in geophysical flows or by laboratory experiments. Therefore, its value in the stellar context remains subject to uncertainties. 
One also has to assume that the temperature gradient in the overshooting layers is everywhere close to the adiabatic value. Finally, one is left with a free parameter that is the number of plumes and one has to assume that all the energy, which is not carried out by radiation, is carried out in the plumes. 
However, this last hypothesis overlooks the fact that the energy generation rate is, with a very good approximation, the same at all points of a spherical surface, as the convective fluctuation of the temperature and the density are very small and one needs to explain how the energy flux would be concentrated into the plumes. Moreover, this assumption cannot be maintained as soon as the equations averaged over a spherical surface are taken into account. In addition, it is also fundamental to consider the kinetic energy flux because, as will be shown without any assumption in this article, if it is neglected, overshoot is impossible. We must also take into account that this flux has different signs in the up- and down-flows (while the convective flux has the same sign in both regions) and that it is non-negligible at every point on a spherical surface. This holds even in the local mixing-length theory (hereafter LMLT) for which only its average value vanishes. Therefore, an extra equation, which  enables us to compute the super-adiabatic gradient at every point of the convective region, must be considered. It allows us to remove the adiabatic hypothesis in the overshooting layers.

Consequently, we will take into account the equations of continuity, mechanical energy conservation, and  thermal energy conservation integrated over the surface of a sphere. This set of equations will be supplemented by the same equations but integrated over the “horizontal” section of a plume.  Using the above-mentioned equations enables the system to be closed and it is no longer necessary to use Taylor’s hypothesis. We have, however, to make an approximation, present in all the theories of convection, because we do not know the turbulent energy spectrum: we have to give a formula for the kinetic energy dissipation rate, which is a phenomenon taking place at the end of the turbulent spectrum (i.e. at high wave numbers) using variables characteristic of the low wave numbers. Therefore, we must choose an expression for the kinetic energy dissipation rate that is only an order of magnitude and that introduces a parameter similar to the mixing length.
As a consequence, the theory we propose is essentially a fully coherent non-local mixing-length theory, which offers a self-consistent method to compute the structure of the overshooting layers. Unfortunately, the price to pay is that we are left with an eight-order system (plus essentially two algebraic equations). 
We also note that recent results obtained by 3D simulations of deep stellar convection \citep{Arnett2009} were able to justify the formula giving the kinetic energy dissipation rate that has been used for many years for the above mentioned reason. However, they show that the mixing length is not the pressure scale height, as usually assumed, but the size of the larger eddies, which is close to the size of the convection zone \cite[see also Sect. 2 in][]{Arnett2015}. 

The theory we will develop can in principle be applied to any convective region, however in the outermost layers of convective envelopes, where the temperature gradient is significantly non-adiabatic and where the convective cells observed at the surface are formed, the plume theory is no longer applicable. Therefore to be able to apply the plume theory to stellar envelopes, it will be necessary to use numerical simulations as initial conditions on top of the quasi-adiabatic region. Alternatively, it can be applied if the solutions become quickly independent of the upper boundary, as suggested by \cite{RieutordZahn1995} for a somewhat idealized problem. On the other hand, nothing seems to be in contradiction with the application of the theory in convective cores. This motivated our choice to develop the theory for a convective core with gas moving upwards in the plumes. Nevertheless, we will indicate how easy it is to modify the equations for applications to envelopes. 

Let us recall that a convective zone boundary is the point where both the radial component of the velocity is cancelled out ($V_r=0$) and radiation carries out all the energy ($L=L_R$, where $L$ and $L_R$ are the total and radiative luminosities, respectively). We now still have to define precisely what the overshooting region is. We would like to define it as the layers of the convective core laying above the core boundary predicted by the LMLT. The boundary given by the LMLT satisfies the two  conditions $V_r=0$ and $L=L_R$. It is worth recalling that it is mathematically not allowed to find that boundary using an interpolation of either $\nabla_R-\nabla_a$ (Schwarzschild criterion, where $\nabla_R$ and $\nabla_a$ are the radiative and adiabatic gradients, respectively) or $\nabla_R-\nabla_L$ \citep[this is the Ledoux criterion, see][their Eq. (10), for a definition of $\nabla_L$]{Gabriel2014} when that function is discontinuous at the surface of the core \citep[see][]{Gabriel2014}. When a better theory is used, we see that the conditions $L=L_R$ and $\nabla_a=\nabla_R$ are not fulfilled at the same point and therefore these two conditions are not equivalent to define the top of the convective core without overshooting. We will thus take the bottom of the overshooting layers at the point where  $L=L_R$ and $V_r \neq 0,$ as this is the condition that is physically meaningful while the other one,   $\nabla_a=\nabla_R$, is equivalent only in the frame of the LMLT. 
\cite{Meakin2007a} give another definition of the convection zone boundaries. They are given by the points where the bulk Richardson number becomes larger than $1/4$. Even if there is some uncertainty over what expression of that number should be used \cite[e.g.][]{Meakin2007a,Arnett2015,Cristini2017}, when the convective core is receding our definition and theirs are practically equivalent. The rise of the potential wall produced by the molecular weight gradient, which gives rise to a fast increase of the Brunt-Väisälä, will quickly stop raising matter already strongly slowed down in the layers with a sub-adiabatic temperature gradient, which  starts from below the overshooting region as we define it. In contrast, the definition used by \cite{Meakin2007a} will be very useful when the convection core expends, all the more so as they give an expression for the entrainment velocity. We will come back to this point in Sect.~\ref{section_8} devoted to expanding cores. 

The paper is organized as follows: Sect.~\ref{section_2} introduces the basic equations of the problem while Sects.~\ref{mean_field} and \ref{section_4} discuss the mean-field and plume-averaged equations, respectively. The system of equations to be integrated is summarized in Sect.~\ref{section_5} while a discussion on the treatment of parameters is provided in Sect.~\ref{section_6} and the boundary conditions in Sect.~\ref{section_7}. In Sect.~\ref{section_8}, the case of an expanding convective core is discussed. Section \ref{section_9} is dedicated to the conclusions. 

\section{Equations of the problem}
\label{section_2}

In the following, we will  make use of  two usual hypotheses for convection, namely: stationarity and vanishing pressure fluctuations as justified in \cite{MassaguerZahn1980}. These two simplifying assumptions are indeed open to criticism. First, the hypothesis of vanishing pressure fluctuations is not sustained by numerical simulations. Second, as could be expected, the hypothesis of stationarity contradicts a basic characteristic of  convection. At its surface, waves are irregularly produced with time in the stable region causing extra mixing. \cite{Arnett2011}  discovered that during late phases of stellar evolution numerical simulations present spatial and time variations in the velocity field, which lead to oscillations similar to those observed in irregular and semi irregular variables. These oscillations caused by luminosity fluctuations directly associated with turbulent convective cells are inherently nonlinear, that is to say they are not detectable through the usual linear studies, and are produced by what they call the $\tau$-mechanism \citep[see also][]{Smith2014}.  Also, our theory predicts one kind of plume, which represents the average of all plumes, but strong plumes will lead to more mixing and this mixing is irreversible. 

Moreover, we will consider chemically homogeneous convective cores. This is justified for a shrinking convective core as encountered in massive main-sequence stars. Indeed, for those stars the core mass grows quickly enough so that when it reaches its maximum value, the star is still chemically homogeneous with a good approximation. In that case, the total mass of the core (including the overshoot region) decreases during the evolution as does the core mass obtained from the LMLT. 
The case of an expanding convective core with a discontinuity of chemical composition at its surface will be briefly considered in Sect.~\ref{section_8}. 

The equations as given in this section are taken from \cite{LedouxWalraven1958} in which any variable $y$ is split into two parts   
\begin{align}
\label{def_average}
y=\overline{y}+\Delta y \;  ,
\end{align}
where $\overline{y}$ is the mean value of $y$, averaged over a spherical surface.  We will consider the variables $y=\{p,T,\rho,\rho V_i\}$, where $p$ is the pressure, $T$ the temperature, $\rho$ the density, and  $\rho V_i$ the mass flux ($\vec V$ being the convective velocity) with $\overline{\rho V_i} = 0$. Since we consider a hypothetic stationary convection, we have here to consider spatial averages only and we do not, as in the more realistic numerical simulations, have to distinguish between spatial and time averages. 

\subsection{Equation of state}

As already mentioned, the pressure fluctuations are neglected $(\Delta p = 0)$ and the  convective zone is considered to be chemically homogeneous. Thus, one has 
\begin{align}
\label{def_state}
\left(\derivp{\ln p}{\ln T}\right)_\rho \frac{\Delta T}{\overline{T}} + \left(\derivp{\ln p}{\ln \rho}\right)_T \frac{\Delta \rho}{\overline{\rho}} = 0 \; , 
\end{align}
or equivalently
\begin{align}
\frac{\Delta \rho}{\overline{\rho}} + Q \frac{\Delta T}{\overline{T}} = 0 \quad {\rm with} \quad 
Q = - \left(\derivp{\ln \rho}{\ln T}\right)_p. \; 
\end{align}
Rigorously, instead of Eq.~(\ref{def_state}) one should write
\begin{align}
\left(\derivp{\ln p}{\ln T}\right)_\rho \frac{\Delta T}{\overline{T}} + \left(\derivp{\ln p}{\ln \rho}\right)_T \frac{\Delta \rho}{\overline{\rho}} + \sum_{i} \left(\derivp{\ln p}{\ln X_i}\right)_{\rho,T} \frac{\Delta X_i}{\overline{X_i}} = 0 \; , 
\end{align}
where $X_i$ is the fraction by weight of nuclei of type $i$. One should also add 
 to the kinetic equations for nuclear reactions a turbulent diffusion term (which is not well known). 
This would make the problem much more complex while introducing a non-negligible effect only in a very tiny layer at the surface of the core where the lifetime of the convective elements becomes of the same order as the thermal diffusion time. Therefore, we will ignore it and consider only \eq{def_state}.

Let us also recall the well-known formula for the adiabatic gradient \citep[see for instance][Eq. 16.3]{Landau1967}
\begin{align}
\nabla_{a} = - \frac{\overline{p}}{C_p \overline{\rho} \overline{T}} \left(\derivp{\ln \rho}{\ln T}\right)_p = 
\frac{Q \overline{p} }{C_p \overline{\rho} \overline{T}}\, ,
\end{align}
where $C_p$ is the heat capacity at constant pressure. 

\subsection{Continuity equation}

For stationary convection, if we assume as is usually done that $V_\phi$ is independent of $\phi$, the mass conservation equation can be written
\begin{align}
\label{mass_conservation}
\frac{1}{r^2} \derivp{}{r} \left(r^2 \rho V_r\right) + \frac{1}{r \sin \theta} \derivp{}{\theta} \left(\sin \theta \rho V_\theta\right) = 0 \; , 
\end{align}
where $V_r,V_\theta,V_\phi$ are the three components of the convective velocity ($\vec V$). 

From \eq{mass_conservation}, it follows that  there is no net mass flux over a spherical surface, that is, that \emph{}  $\overline{\rho \vec V} = 0$.  Therefore, one has 
\begin{align}
\label{averaged_continuity}
\overline{\rho} \, \overline{\vec V} + \overline{\Delta \rho \, \vec V} = 0 \, .
\end{align}
As $\overline{\rho \vec V} = 0$, we prefer to use $\rho \vec V$ as a fundamental variable rather than $\vec V$.

\subsection{Mechanical energy conservation}

The equation of motion, assuming $\Delta p=0$, is
\begin{align}
\label{momentum_conservation}
\vec \nabla \cdot \left( \rho \vec V \vec V \right) = \rho \vec V \cdot \left(\vec \nabla \cdot \vec V\right) = \frac{\Delta \rho}{\rho} \vec \nabla \overline{p} - \alpha \frac{\rho \vec V \left \vert V_r \right \vert}{l_{\rm MLT}} \; .
\end{align} 
The coefficient $\alpha$ as dissipation factor was first introduced by Prandl \citep[see references in][]{Arnett2015} but the same formula was also introduced independently by \cite{Cowling1935}. The last term, which comes from the dissipation by viscosity of kinetic energy at the low end of the energy cascade, is defined by using $1/\tau = \alpha \left\vert V_r \right\vert / l_{LMLT}$  for the characteristic time ($\tau$) of kinetic energy dissipation into heat. Most often, $l_{LMLT}$ is approximated by  $l_{LMLT}\approx H_p$, as has been the case since Böhm-Vitense. In contrast, the recent work by \cite{Arnett2015} suggests the use of $1/\tau =\left\vert V \right\vert / l_{d} $ with $l_d \approx 0.8 l_{CZ}$ (see their Table 1), where $l_{CZ}$ is the thickness of the convection zone. This is an important change as it makes convection much more efficient and reduces the departure from adiabatic gradient. We  momentarily keep the old definition of “the mixing length” to allow for the following discussion, but we will adopt Arnett et al.'s definition from \eq{kinetic_energy_flux} on. 

We also assumed that the static equilibrium equation is given by 
\begin{align}
\deriv{\overline{p}}{r} = -\overline{\rho} g  \; , 
\end{align}
with $p$ the gas pressure. The latter equation is verified to a good accuracy in the deep interior of stars where $\vec V \ll \vec c$ ($\vec c$ being the sound speed). 
        
The last term of \eq{momentum_conservation} represents the divergence of the viscous stress tensor where $l_{\rm MLT}$ is the mixing length. 
The coefficient $\alpha$ is introduced in order to recover the equations of the LMLT for the local solution of the linearized equations for convection \citep[see][for details]{Gabriel1974,Gabriel1975,Gabriel1996}. For instance, to obtain the equation of \cite{Henyey1965}, we must use the value $\alpha=8/3$. However, there is no reason to also adopt this value in a convective core. Nuclear reactions are going on in the core and also the energy losses are not necessarily isotropic as assumed in the outer layers. 

However, to simplify the notations, we have chosen here to set $\alpha$ equal to unity. It is equivalent to define a new mixing length $l=l_{\rm MLT}/\alpha$ and to define the characteric dissipation time of kinetic energy by $\tau = l / \left \vert V_r \right \vert$. This means that here the mixing length must not be interpreted as the mean travel distance of a turbulent eddy but as the length that allows us to define the correct characteristic time $\tau$. It was already so in the formalism adopted by \cite{Gabriel1974,Gabriel1975} and \cite{Gabriel1996}.

To go further, we multiply \eq{momentum_conservation} by $\vec V$ to get the equation of kinetic energy conservation
\begin{align}
\label{kinetic_energy_flux}
\vec \nabla \cdot \left(\frac{1}{2} \rho V^2 \vec V\right) =  \frac{\Delta \rho}{\rho} \vec V\cdot \vec \nabla \overline{p} -  \frac{\rho \vec V^2 \left \vert V_r \right \vert}{l} \, , 
\end{align}
where we introduced our above-mentioned definition of $l$. The parameter $l$ used in the above equation is related to that defined by \cite{Arnett2009,Arnett2015} by 
\begin{align}
l=l_d \frac{\left \vert V_r \right \vert}{\left \vert V \right \vert} \, . 
\end{align}
They also show that this expression for the kinetic energy dissipation is obtained because that term has the Kolmogorov form. 

  We also introduce the kinetic energy flux of turbulence  $\vec F_K=\rho V^2 \vec V / 2$ and the expression adopted for the kinetic energy dissipation rate into heat
$ \rho \epsilon_2=  \rho V^2 / \tau = \rho V^2 \left \vert V_r \right \vert / l$ .  
It provides for the linearized equations of convection a local stationary solution identical to the LMLT ones \citep[see][]{Gabriel1974,Gabriel1975,Gabriel1996}. 

Equation~(\ref{kinetic_energy_flux}) averaged over a spherical surface then gives
\begin{align}
\label{mechanical_conservation_averaged}
\vec \nabla \cdot \overline{\vec F_K} = \overline{\frac{\Delta \rho}{\rho} \vec V} \cdot \vec \nabla \overline{p} 
- \overline{\rho \epsilon_2} \; .
\end{align}

\subsection{Thermal energy conservation}

Let us now consider the thermal energy equation
\begin{align}
\label{energy_conservation_eq1}
\vec \nabla \cdot \left(  \vec F_H + \vec F_R\right) - \vec V \cdot \vec \nabla p = \rho \epsilon + \frac{\rho \vec V^2 \left \vert V_r \right \vert}{l} \; ,
\end{align}
where $F_R$ is the radiative flux, $\epsilon$ is the energy generation rate, and $F_H$ is the enthalpy flux defined as
\begin{align}
\vec F_H \equiv \rho U \vec V + p \vec V 
= \left( \overline{U} + \frac{\overline{p}}{\overline{\rho}} + \overline{T} \Delta S \right) \rho \vec V 
\end{align}
with $S$ being the entropy. Using \eq{mass_conservation}, the divergence of the enthalpy flux can be written as
\begin{align}
\label{energy_conservation_eq2}
 \vec \nabla \cdot \vec F_H = \rho \vec V \cdot \left(\frac{\vec \nabla \overline{p}}{\overline{\rho}} + \overline{T} \vec \nabla \overline{S}  \right) + \vec \nabla \cdot \vec F_C, 
\end{align}
where the convective flux, $F_C$, is defined by
\begin{align}
\label{def_convective_flux}
\vec F_C \equiv \overline{T} \Delta S \rho \vec V \; .
\end{align}
The isentropic assumption has not been made so that \eq{energy_conservation_eq2} can be considered as a generalization of the energy equation used in \cite{RieutordZahn1995}. 

Using Eqs.~(\ref{kinetic_energy_flux}) and (\ref{energy_conservation_eq2}), \eq{energy_conservation_eq1} finally becomes
\begin{align}
\label{equation_fluxes}
 \vec \nabla \cdot \left(  \vec F_K + \vec F_C + \vec F_R\right) 
 + \rho \vec V \cdot \left(  \overline{T} \vec \nabla \overline{S}  \right)  = \rho \epsilon. 
\end{align}
 Averaging over a spherical surface, we recover the usual equation of thermal energy conservation
    \begin{align}
    \label{equation_mean_fluxes}
 \vec \nabla \cdot \left(  \overline{\vec F_K} + \overline{\vec F_C} + \overline{\vec F_R} \right) 
   = \overline{\rho \epsilon} 
 \end{align}
 or equivalently 
\begin{align}
\label{conservation_luminosity}
\overline{L_K} + \overline{L_C}+\overline{L_R} = \overline{L,}
\end{align}
where $\overline{L}_K,\overline{L}_C,\overline{L}_R$ are the average of the  kinetic, convective,  radiative luminosities, and $\overline{L}$ the total luminosity. 
By taking the pressure fluctuations into account and by introducing a new term that implies that energy generation also supplies the work needed to maintain the convective flow, \cite{Arnett2009} (Sect. 3) and \cite{Meakin2007a} (Appendix A) obtained a different expression for \eq{equation_fluxes}. Also if the driving of turbulence may be local, turbulent flows redistribute turbulence more uniformly, implying non-zero turbulent flux. 

At this point, it is worth mentioning that convection is more efficient in rising plumes rather than in the downward flow because the convective flux is larger (except if the surface area occupied by up- and down-flows is the same) and also because of the positive kinetic energy flux (while it is negative in the down-moving gas). Therefore, rising plumes carry out more energy than produced inside the plumes since $\epsilon$ is, with a good approximation, the same everywhere on a spherical surface. For instance, if in a thin layer the contribution of the radial component of the fluxes to $\vec \nabla \cdot \left(\vec F_K+\vec F_C \right)$ averaged over a plume is larger than its value averaged over a spherical surface, the difference must be counterbalanced by the contribution of the horizontal components directed inwards to the plume to prevent the gas of that layer from cooling down. This horizontal flux is necessary to allow a stationary state. This remark holds also for the LMLT even if, in that case, the flux of kinetic energy from the down to the up-flow allows a kinetic energy flow averaged over a spherical surface equals to zero.

Subtracting \eq{equation_mean_fluxes} from \eq{equation_fluxes} we further get 
\begin{align}
\label{eq_thermal_div_fluxes1}
\vec \nabla \cdot \left(  \Delta \vec F_C  + \Delta \vec F_K + \Delta \vec F_R\right) + \rho \vec V \cdot \left(  \overline{T} \vec \nabla \overline{S} \right) = \Delta \left(\rho \epsilon\right) \, .
\end{align}

In the framework of the LMLT, the first term of the latter equation reads
\begin{align}
\vec \nabla \cdot \left(\Delta \vec F_C \right) = \frac{\rho T \Delta S \left\vert V_r \right\vert }{l_{\rm MLT}} \, ,
\end{align}
and the kinetic energy flux ($\vec F_K$) is ignored. However, $\vec F_K$ is everywhere different from zero and only its average over a spherical surface  vanishes. Therefore, it is not justified to assume that $\Delta \vec F_K = 0$ and we have to add an extra term. To this end, we will assume that 
\begin{align}
\label{eq_thermal_div_fluxes2}
\vec \nabla \cdot \left( \Delta \vec F_K + \Delta \vec F_C \right) = \frac{\rho T \Delta S \left\vert V_r \right\vert }{\alpha l} + \alpha_2 \frac{k}{2} \frac{\rho V_r^3}{l} \; .
\end{align}
The computation of the fluctuation of the convective and kinetic energy fluxes would normally require knowledge of higher-order fluctuations of  the temperature and of the impulsion. Therefore, \eq{eq_thermal_div_fluxes2} may be considered as the second closure hypothesis of the problem, the first one being the expression chosen for $\epsilon_2$. Assuming that the turbulent cascade follows the Kolmogorov law proportional to the third power of the velocity, \cite{Arnett2009} have justified this expression for $\epsilon_2$. 
Also numerical simulations do not need any closure equation as departures from averages are obtained through the simulations and all average quantities required, for instance, in Eqs. (20) to (24) of Arnett et al (2009), can be obtained through time or spatial averages. 

Within the LMLT, the first term of the right hand side of \eq{eq_thermal_div_fluxes2} is introduced because, after having travelled over an average mixing length, a rising element releases its excess thermal energy into the average medium. Here, it will be interpreted as the dissipation rate of thermal energy. We add a second term in \eq{eq_thermal_div_fluxes2} because the kinetic energy must also be released into the average medium. Here, this term gives the dissipation rate of convective kinetic energy. 
When convection at one point is compared with the average over a spherical surface, the last term of \eq{eq_thermal_div_fluxes2} must be present. In contrast, in the LMLT, this term is absent. This can be explained as follows; the LMLT has been developed for outer layers, for which the two right-hand-side terms of \eq{eq_thermal_div_fluxes2} are of the same order of magnitude.   These terms are only orders of magnitudes so that they can be gathered together into the first one and it may be considered that the LMLT takes both terms into account.
 
 Finally, using \eq{eq_thermal_div_fluxes1} together with \eq{eq_thermal_div_fluxes2}, one gets 
 \begin{align}
 \label{eq_themaldiv_fluxes3}
 \vec \nabla \cdot \left(\Delta \vec F_R\right) + \rho \vec V \cdot \left(\overline{T} \vec \nabla \overline{S}\right) 
 = \Delta \left(\rho \epsilon\right) - \frac{\rho T \Delta S \left\vert V_r \right\vert}{\alpha l} 
 - \alpha_2 \frac{k}{2} \frac{\rho V_r^3}{l} \, , 
 \end{align}
 with
\begin{align}
\label{delta_rho_epsilon}
\Delta \left(\rho \epsilon\right)  
=  \overline{\rho \epsilon} \left[ \nu - \left(Q+1\right) \mu \right] \frac{\Delta T}{\overline{T}} \, , 
\end{align} 
where we used the definitions $\nu \equiv  \left(\derivp{\ln \epsilon}{\ln T}\right)_\rho$ and $\mu \equiv  \left(\derivp{\ln \epsilon}{\ln \rho}\right)_T$. 

\section{Mean equations}
\label{mean_field}

\subsection{Horizontal dependence of the convective variables}
\label{defs_horizontal}

To go further, we now have to specify the horizontal dependence of the convective variables. More precisely, we will separate the 
 variables in the following way:
\begin{align}
\label{defs_horizontale}
&\rho \left(r,\theta,\phi\right) = \overline{\rho(r)} + \Delta \rho \left(r, \theta, \phi \right) = \overline{\rho(r)} + \Delta \rho_0(r) \, h \left( \theta, \phi\right) \, , \nonumber \\
&T \left(r,\theta,\phi\right) = \overline{T(r)} + \Delta T \left(r, \theta, \phi \right) = \overline{T(r)} + \Delta T_0(r) \, h \left( \theta, \phi\right) \, , 
\end{align}
where $\overline{\rho(r)}$ and $\overline{T(r)}$ are averages over a spherical surface as defined by \eq{def_average}. 

For the mass flux, as already mentioned, its average vanishes so that 
\begin{align}
\label{def_mass_flux}
\rho \vec V \left(r,\theta,\phi\right) =  (\rho \vec V )_0(r) \, h \left( \theta, \phi\right)  \, , 
\end{align}
where $(\rho \vec V)_0$ is the value of $\rho \vec V$ at the point where $h$ is chosen to be equal to unity. This holds also for $\Delta \rho_0$ and $\Delta T_0$.

The normalization of the horizontal function, $h$, gives 
\begin{align}
\label{normalization_h}
\int_S h \left( \theta, \phi\right) {\rm d}\Omega = 0 \, , 
\end{align}
where ${\rm d}\Omega = \sin \theta \, {\rm d}\theta \, {\rm d}\phi$ is the elementary solid angle, and $S$ the surface of a sphere. Equation~(\ref{normalization_h}) ensures that the convective flux of matter averaged over a spherical surface is zero.

For the convective velocity, we must separate the variables with another horizontal function because the average of the velocity does not vanish. Therefore, we write 
\begin{align}
\label{def_velocity}
\vec V = \vec V_0 (r) \, h_1 \left( \theta, \phi\right) \, . 
\end{align}
The functions $h$ and $h_1$ can be provided by numerical simulations. Moreover, using Reynold-averaged Navier-Stokes equations (RANS), there is no need to provide such a decomposition and subsequently to make a choice concerning their formulation (see below). 

Using a Taylor expansion at first order, one can also write for the convective velocity 
\begin{align}
\label{velocity_taylor}
\vec V \simeq \frac{\rho \vec V}{\overline{\rho}} \, \left(1-\frac{\Delta \rho}{\overline{\rho}}\right) 
= \frac{\left(\rho \vec V\right)_0}{\overline{\rho}} \, h \left[ 1 - \left(\frac{\Delta \rho_0}{\overline{\rho}}\right) \, h \right] \, .
\end{align}
In convective cores, the Mach number (defined as the ratio between the convective velocity and the sound speed) is very small. Therefore, $\Delta \rho / \overline{\rho} \ll 1$, because it depends on the squared Mach number. Consequently, equating \eq{def_velocity} and \eq{velocity_taylor}, one has 
\begin{align}
h_1 \simeq h \quad {\rm and} \quad \overline{\rho} \vec V_0 \simeq \left(\rho \vec V\right)_0 \, .
\end{align}

It is now possible to express the averaged convective quantities. Let us first consider the averaged convective flux. Using \eq{def_mass_flux}, it reads 
\begin{align}
\label{averaged_convective_flux}
\overline{\vec F_C} &= C_p \overline{\Delta T \rho \vec V} 
= C_p \Delta T_0 \left(\rho \vec V\right)_0  G_1 \, ,
\end{align}
where we have defined 
\begin{align}
G_1 = \frac{1}{S} \int_S h^2 r^2 {\rm d}\Omega \, .
\end{align}

For the convective velocity, using \eq{averaged_continuity} together with \eq{def_mass_flux}, we have 
\begin{align}
\overline{\vec V} = - \frac{\Delta \rho_0(r)}{\overline{\rho}} \vec V_0 (r) \frac{1}{S} \int_S h h_1 r^2 {\rm d}\Omega 
= - \frac{\Delta \rho_0(r)}{\overline{\rho}} \vec V_0 (r) \, G_2 \, , 
\end{align}
with 
\begin{align}
G_2 = \frac{1}{S} \int_S h h_1 r^2 {\rm d}\Omega \simeq \frac{1}{S} \int_S h^2 r^2 {\rm d}\Omega = G_1\, .
\end{align}

It is now worth considering the averaged kinetic energy flux and more precisely its radial component. One can write 
\begin{align}
\label{averaged_kinetic_isotropy}
\overline{F}_{K,r} &= \overline{\frac{1}{2} \rho V^2 V_r} = \frac{k}{2} \overline{\rho V_r^3} \, .  
\end{align}
To obtain \eq{averaged_kinetic_isotropy} we have introduced a hypothesis concerning the isotropy of convection and we have written
\begin{align}
\label{isotropic_parameter}
V^2 = k V_r^2, 
\end{align}
with $k=3$ for isotropic convection and $k=1$ for purely radial convection. This is a significant advantage of using the equation of kinetic energy conservation rather than the radial component of the equation of motion as the latter takes a much more complex form when convection is anisotropic so that the previous  works implicitly assume isotropy \citep[e.g.][]{RieutordZahn1995,LoSchatzman1997}. Of course, we introduce a new free parameter but at least it will be possible to check the influence of the choice of its value on the results while previous works were confined to k=3.  \cite{Arnett2009} discussed the anisotropy far away from the boundaries (see their Eq. (10) in Sect. 2.4) and found that $1.5< k < 2.28$. 

Using \eq{averaged_kinetic_isotropy}, the averaged kinetic energy flux immediately reads 
\begin{align}
\label{averaged_kinetic_luminosity}
\overline{F}_{K,r} &= \overline{\frac{1}{2} \rho V^2 V_r} = \frac{k}{2} \overline{\rho V_r^3} \nonumber \\
&= \frac{k}{2} \left(\rho V_r^3\right)_0 \frac{1}{S} \int h h_1^2 r^2 {\rm d}\Omega =  \frac{k}{2} \left(\rho V_r^3\right)_0 G_3 \, ,
\end{align}
where we have introduced the notation
\begin{align}
\left(\rho V_r^3\right)_0 \equiv \left(\rho V_r\right)_0 V_{0,r}^2 \,  
\end{align}
and
\begin{align}
G_3 = \frac{1}{S} \int_S h h_1^2 r^2 {\rm d}\Omega \simeq \frac{1}{S} \int_S h^3 r^2 {\rm d}\Omega \, .
\end{align}
The averaged luminosity associated with the radial component of the averaged kinetic energy flux can then be written as $\overline{L}_K = 4\pi r^2 \overline{F}_{K,r}$. 

Finally, one can express the kinetic energy dissipation rate into heat using the isotropic parameter (Eq.~\ref{isotropic_parameter}), such as
\begin{align}
\overline{\rho \epsilon_2} = \frac{ \overline{\rho V^2 \vert V_r \vert}}{l} = \frac{ k \overline{\rho V_r^2 \vert V_r \vert}}{l} \, .
\end{align}
Therefore, one has
\begin{align}
\label{rho_epsilon_2}
\overline{\rho \epsilon_2} = \frac{ k (\rho V_r^3)_0}{l} \frac{1}{S} \int_S h h_1 \left\vert h_1 \right\vert r^2 {\rm d}\Omega 
= \frac{ k (\rho V_r^3)_0}{l} \, G_4 \, , 
\end{align}
with 
\begin{align}
G_4 =  \frac{1}{S} \int_S h h_1 \left\vert h_1 \right\vert r^2 {\rm d}\Omega \simeq \frac{1}{S} \int_S h^2 \left\vert h \right\vert   r^2 {\rm d}\Omega \, .
\end{align}

To be able to compute the coefficients $G_i$ and in the following section the equations integrated over one plume, we must give an explicit form to the angular function $h(\theta,\phi)$. We will first assume that there are $N$ identical plumes, which have a conical shape of  opening $\theta_M (r)$. Moreover, within each plume, we assume
\begin{align}
\label{def_h}
h(\theta,\phi) \equiv \cos \left(\frac{\pi}{2} \frac{\theta}{\theta_M}\right) \,.
\end{align}
This means that we take $h=1$ on the axis of the cone associated with each plume. Equation (\ref{def_h}) is different from the usual one,  which  assumes a Gaussian profile that fits well the laboratory observations \citep[e.g.][]{Schmitt1984,RieutordZahn1995,LoSchatzman1997}. We have preferred it because it clearly defines the boundary for a plume, which is necessary when the descending regions are also taken into account. If we want to consider descending plumes as encountered in stellar envelopes, we just have to multiply the expressions given above for the convective variables by $-1$ or take $h=-1$ on the axis of the cone. 

Therefore, using \eq{def_h}, the convective fluctuations of density and temperature read 
\begin{align}
\frac{\Delta \rho}{\overline{\rho}} &= \frac{\Delta \rho_0}{\overline{\rho}} \cos \left(\frac{\pi}{2} \frac{\theta}{\theta_M}\right) \\
\frac{\Delta T}{\overline{T}} &= \frac{\Delta T_0}{\overline{T}} \cos \left(\frac{\pi}{2} \frac{\theta}{\theta_M}\right) \, ,
\end{align}
and for the radial mass flux 
\begin{align}
\rho V_r  &=  (\rho V_r )_0 \, \cos \left(\frac{\pi}{2} \frac{\theta}{\theta_M}\right)  \, . 
\end{align}
Only the equation of mechanical energy conservation is used in this work instead of the radial component of the equation of motion as done in the above-mentioned works on plumes. Consequently, only the value of $\rho V_\theta$ at the boundary of a plume will appear in the equations and we do not have to make an assumption on its angular variation. 

We also have to introduce an expression for $h(\theta,\phi)$ outside the ascending plumes, where matter moves downwards everywhere.  
To do so, we will consider two hypotheses; 
\begin{itemize}
        \item First, we assume that there are $N_D$ descending regions that have the same geometry as the plumes. Therefore, 
        \begin{align}
        \label{h_down_hyp1}
         h(\theta,\phi)= A_D \cos \left(\frac{\pi}{2} \frac{\theta}{\theta_{M,D}}\right) \, ,
        \end{align} 
        where the coefficient $A_D$ will be deduced from the mass conservation. 
        \item Second, we assume that each plume is surrounded by a hollow cone with an opening between $\theta_M$  and $\theta_{M,D}$, where the flow is directed downwards. In this case, one has
        \begin{align}
        \label{h_down_hyp2}
        h(\theta,\phi)= A_D \sin \left(\pi\frac{\theta-\theta_M}{\theta_{M,D}-\theta_M}\right) \, 
        \end{align}  
    and $N_D=N$.
\end{itemize}
We will have to assume that the sum of the fraction of a spherical surface occupied by plumes and by descending gas equals to unity (i.e., $S_M+S_D=1$). Obviously, this hypothesis is an approximation as it is impossible to cover completely a spherical surface with spherical caps. 
Also $S_M=N(1-\cos \theta_M )/2$, while the expression for $S_D$ will be different for the two hypotheses considered (see Appendix~\ref{appendixA}). 

\subsection{Explicit form of the equations averaged over a spherical surface}

In the following, we will consider two cases: that one can use either \eq{h_down_hyp1} or \eq{h_down_hyp2}, depending on the treatment of the downward flow. As already discussed in Sect.~\ref{defs_horizontal}, there is no net mass flux ($\overline{\rho \vec V} = 0$) so that the average of $h$ over a spherical surface vanishes. Thus, using the normalization condition, \eq{normalization_h}, together with \eq{def_h} and \eq{h_down_hyp1}, 
allows us to write 
\begin{align}
\label{mass_flux_AD1}
N J_{0,M} + A_D N_D  J_{0,D} = 0 \, , 
\end{align}
where we have defined 
\begin{align}
J_{0,M}  &= \int_0^{\theta_M} \cos \left(\frac{\pi}{2} \frac{\theta}{\theta_M} \right) \sin \theta \, {\rm d}\theta \nonumber \\
 J_{0,D} &= \int_0^{\theta_{M,D}} \cos \left(\frac{\pi}{2} \frac{\theta}{\theta_{M,D}} \right) \sin \theta \, {\rm d}\theta \, .
\end{align}
Alternatively, if we use \eq{h_down_hyp2} instead of \eq{h_down_hyp1}, one gets 
\begin{align}
\label{mass_flux_AD2}
 J_{0,M} + A_D  J^\prime_{0,D} = 0 \, , 
\end{align}
where 
\begin{align}
J_{0,D}^\prime &= \int_{\theta_{M}} ^{\theta_{M,D}} \sin \left(\pi\frac{\theta-\theta_M}{\theta_{M,D}-\theta_M}\right)  \sin \theta \, {\rm d}\theta. 
\end{align}
Therefore, the conservation of the mass flux, through \eq{mass_flux_AD1} and \eq{mass_flux_AD2}, provides the coefficient $A_D$. 
The integrals $J_{i,M}, J_{i,D}$ , and $J_{i,D}^\prime$ are given in Appendix~\ref{appendixA}. 

To proceed further, we must consider that numerical simulations show, for convective envelopes, that the physical properties of convection are very different inside and outside the plumes \cite[e.g.][]{Stein1998}. To take this into account, we must consider that the anisotropy parameter ($k$) may have different values inside and outside the plumes. Therefore, we will consider the parameter $k$ in the plumes and $k_D$ in the downward flow. 
Indeed, one would expect that convection is more isotropic outside the plumes and therefore that $k$ is closer to three than inside them. This is what numerical simulations for the solar envelope show (R. Samadi, private communication). They also show that convection is nowhere isotropic; this shows the advantage of using the equation of mechanical energy conservation rather than the equation of motion.  
It is also possible that the mixing length has different values ($l$ and $l_D$) in the up- and down-moving flows. We will thus formally consider $l_D\neq l$ in the following. However, as the ratio $l_D/l$ is unknown, a pragmatic approach would be to take $l_D=l$ in order to avoid multiplying the number of free parameters. 
Also \cite{Arnett2009} and \cite{Viallet2013} showed from their numerical simulations that we can take $l=l_D\approx l_{CZ}$.

Using the same procedure, \eq{mechanical_conservation_averaged}, together with \eq{averaged_kinetic_luminosity} and \eq{rho_epsilon_2}, gives for the conservation of mechanical energy 
\begin{align}
\label{mean_mechanical}
\frac{1}{r^2} \deriv{}{r} \left( \frac{k r^2 }{2} \left(\rho V_r^3\right)_0 G_3\right) &= 
-\left(\frac{\Delta \rho_0}{\overline{\rho}}\right) \left(\rho V_r\right)_0 g \,  G_1 \nonumber \\
&- \frac{ k \left(\rho V_r^3\right)_0}{l} \, G_4 \, , 
\end{align}
where, using \eq{h_down_hyp1}, the angular integrals $G_i$ are given by 
\begin{align}
\label{defs_Gi}
G_1 &= G_2 = \frac{N}{2} J_{1,M} + \frac{N_D A_D^2}{2} J_{1,D}\nonumber \\
G_3 &= \frac{N}{2} J_{4,M} + \frac{k_D}{k} \frac{N_D A_D^3}{2} J_{4,D} \nonumber \\
G_4 &= \frac{N}{2} J_{4,M} + \frac{l\, k_D}{l_D\, k} \frac{N_D \left\vert A_D^3 \right \vert}{2} J_{4,D} \, , 
\end{align}
and if \eq{h_down_hyp2} is used instead of \eq{h_down_hyp1},  the integrals $J_{i,D}$ have to be replaced by $J_{i,D}^\prime$. 

For the conservation of the total luminosity (Eq.~\ref{conservation_luminosity}), using \eq{averaged_kinetic_luminosity} and \eq{averaged_convective_flux}, we immediately get
\begin{align}
\label{mean_luminosity}
\frac{\overline{L}(r)-\overline{L}_R(r)}{4\pi r^2} = C_p \Delta T_0 \left(\rho V_r\right)_0 G_1 
+ \frac{k}{2} \left(\rho V_r^3\right)_0 G_3 \, .
\end{align}

It is interesting to note that  if $\overline{F}_{K,r} >0$ (i.e. if the last term of \eq{mean_luminosity} is positive) in the overshooting layers, we are guaranteed to find a solution where  $\overline{F}_{K,r} =0$ or $V_r=0$ as shown in Appendix~\ref{appendixB}. In the opposite case, i.e.  $\overline{F}_{K,r} <0$ in the overshooting layers, it is much more problematic except if the overshooting is very small. 

Although \cite{Arnett2009,Viallet2013}, and \cite{Arnett2015} have shown that the averaged kinetic energy flux cannot be null everywhere ($\overline{F}_{K,r} = 0$), it is interesting to discuss this extreme case because it conforms with the LMLT and corresponds to the case $S_M=S_D=1/2$ when $\theta_M$ is small. Then, \eq{mean_mechanical} and \eq{mean_luminosity} become
\begin{align}
\label{sec3_eq55}
\frac{\Delta \rho_0}{\overline{\rho}} \left(\rho V_r\right)_0 g G_1 
+ \frac{k \left(\rho V_r^3\right)_0}{l} G_4 = 0 \, , 
\end{align}
and 
\begin{align}
\frac{\overline{L}(r)-\overline{L}_R(r)}{4\pi r^2} = C_p \Delta T_0 \left(\rho V_r\right)_0 G_1 \, .
\end{align}
The second term of  \eq{sec3_eq55} is always positive and then $\Delta \rho_0 \left(\rho V_r\right)_0 / \overline{\rho}$  must be negative. Consequently, in the second equation it is impossible to have $\overline{L}(r)-\overline{L}_R (r)<0$,  which means that overshooting is impossible above the core boundary predicted by the LMLT. One can reach the same conclusion using the fundamental equation,  \eq{mechanical_conservation_averaged}, which does not make any assumption concerning the dissipation rate of kinetic energy: 
\begin{align}
\label{kinetic_tmp}
\vec \nabla \cdot \overline{\vec F_K} = \overline{\frac{\Delta \rho}{\rho} \vec V} \cdot \vec \nabla \overline{p} - \overline{\rho \epsilon_2} \, .
\end{align}
We clearly see the consequence of the hypothesis stating that the averaged kinetic energy flux cancels everywhere. Since by definition $\overline{\rho \epsilon_2} >0$, it is necessary that $\overline{\Delta \rho V_r /\rho}<0$ to fulfil the equation of mechanical energy conservation. This means that $\overline{\Delta T V_r / T}>0$, so does $\overline{L}(r)-\overline{L}_R (r)$. Therefore, a non-vanishing turbulent kinetic energy flux is fundamental for the existence of overshooting. This is in opposition to a lot of works that compute overshooting and neglect the kinetic energy flux. They get results because they neglect some basic equations of the problem such as the equation of mechanical energy conservation, and also because they do not consider the downwards moving matter \citep[see][and the works discussed in \citealt{Renzini1987}]{Shaviv1973}. Also, \eq{kinetic_tmp} shows more generally that, in the overshooting layers, the divergence of the kinetic energy flux must be negative and in absolute value larger than the kinetic energy dissipation rate. 

\section{Equation integrated on a spherical section of a plume}
\label{section_4}

We now consider the equations describing plumes. The computation of the integrated expressions is somewhat long but requires only fundamental treatments of the equations. One has just to keep in mind that $\theta_M$ is a function of the radius and also that most terms in the equations are linear in the convective variables with the exception of some terms in Eqs.~(\ref{averaged_continuity}), (\ref{kinetic_energy_flux}),  (\ref{def_convective_flux}), (\ref{eq_thermal_div_fluxes2}), and (\ref{eq_themaldiv_fluxes3}). As long as the order of each term is maintained during the algebraic manipulations, the local value of any variable can be taken as equal to its average over a spherical surface and conversely.

\subsection{Continuity equation}

We start by integrating the equation of mass conservation (Eq.~\ref{mass_conservation}); this gives
\begin{align}
\label{continuity_plume}
&J_{0,M} \deriv{}{r} \left[r^2 \left(\rho V_r\right)_0 \right] 
+ J_{2,M} r^2 \left(\rho V_r\right)_0 \deriv{\ln \theta_M}{r} \nonumber \\
&+ r\sin \theta_M \left(\rho V_\theta\right)_{\theta_M} = 0 \, .
\end{align}

When only the plumes are taken into account, the lack of a physical boundary condition at the edge of the cone forces us to adopt a hypothetical boundary condition and Taylor’s hypothesis is used. However, this closure has been criticized by \cite{Schmitt1984} and \cite{Bonin1996}. The latter pointed out the need to replace Taylor’s hypothesis by a proper closure using the mean-field equations and also 
to take into account the dissipation term in the equation of motion. Here we will not have to use Taylor’s hypothesis as we will use the mean-field equations (see Sect.~\ref{mean_field}). 

\subsection{Mechanical energy conservation}

For the mechanical energy conservation, we use \eq{kinetic_energy_flux}, which after integration gives
\begin{align}
\label{mechanical_plume}
&J_{4,M} \deriv{}{r} \left[r^2\frac{k}{2} \left(\rho V_r^3\right)_0\right] + \frac{3}{2} k r^2 \left(\rho V_{r}^3\right)_0 \deriv{\ln \theta_M}{r} J_{5,M} + r \left(\sin \theta \, F_{K,\theta}\right)_{\theta_M} \nonumber \\
&= - r^2 \left[\frac{\Delta \rho_0}{\overline{\rho}} \left(\rho V_r\right)_0 g J_{1,M} 
+  \frac{k  \left(\rho V_r^3\right)_0}{l} J_{4,M}\right] \, .  
\end{align} 
At the plume boundary $V_r=0$. Therefore, using the hypothesis that the two horizontal components of the convective velocity are equal, we have 
\begin{align}
\left(F_{K,\theta}\right)_{\theta_M} = \left(\rho V_\theta^3\right)_{\theta_M} \, .
\end{align}

It is also worthwhile noticing  that, in a plume, the radial component of the kinetic energy flux is positive while it is negative outside. Moreover, the difference between the values of $\vec \nabla \cdot \vec F_K$ in these regions is due to either the difference of Archimede’s force, or to the difference in the dissipation rate of kinetic energy (through the amplitude term $A_D$), or to the advection of kinetic energy at the plume boundary.

\subsection{Thermal energy conservation}

In this section, we aim at deriving the equation governing the temperature gradient. To this end, we start by integrating \eq{eq_themaldiv_fluxes3} over a horizontal section of a plume. We then obtain
\begin{align}
\label{tmp_thermal_plume}
&\int_{0}^{\theta_M} \left\{\frac{1}{r^2}\derivp{}{r} \left[r^2 \Delta F_{R,r} \right] +
\rho V_r \overline{T} \deriv{\overline{S}}{r} - \Delta \left(\rho \epsilon\right)\right\} r^2 \sin \theta {\rm d}\theta \nonumber \\
&+r^2 \frac{C_p \overline{T}}{\alpha l} \left(\frac{\Delta T_0}{\overline{T}}\right) \left(\rho V_{r}\right)_0 J_{1,M} 
+ \alpha_2 \frac{k}{2} \frac{\left(\rho V_{r}^3\right)_0}{l} r^2 J_{4,M} \nonumber \\
&+ r\left[\sin \theta\left(\Delta F_{R,\theta}\right)\right]_{\theta_M} = 0 \, .
\end{align}

To go further, we have to express the perturbation of the radiative flux. To do so, we place ourselves in the limit of the diffusion and we get for the horizontal component of the radiative flux
\begin{align}
\label{delta_FR_horizontal}
\left(\Delta F_{R,\theta}\right)_{\theta_M} = 
\frac{4ac}{3} \overline{\left(\frac{T^4}{\kappa \rho}\right)} 
\left(\frac{\Delta T_0}{\overline{T}}\right) \frac{\pi}{2} \frac{1}{r \theta_M} \, , 
\end{align}
and for the perturbation of the radial component 
\begin{align}
\label{delta_FR_radial}
\Delta F_{R,r} &= \chi_1 \overline{F}_{R,r} \cos\left(\frac{\pi}{2}\frac{\theta}{\theta_M}\right) + \chi_2 \deriv{}{r} \left[\ln \left(\frac{\Delta T_0}{\overline{T}}\right)\right] \cos\left(\frac{\pi}{2}\frac{\theta}{\theta_M}\right) \nonumber \\
&+ \chi_2 \left(\frac{\pi}{2}\frac{\theta}{\theta_M}\right)  \deriv{\ln \theta_M}{r} \sin \left(\frac{\pi}{2}\frac{\theta}{\theta_M}\right) \, , 
\end{align}
where we defined
\begin{align}
\chi_1 &= \left[(4-\kappa_T)+Q(1+\kappa_\rho)\right]\left(\frac{\Delta T_0}{\overline{T}}\right) \nonumber \\
\chi_2 &= -\frac{4ac}{3} \overline{\left(\frac{T^4}{\kappa \rho}\right)} \left(\frac{\Delta T_0}{\overline{T}}\right) \, .
\end{align}

Finally, using \eq{tmp_thermal_plume} together with Eqs.~(\ref{delta_rho_epsilon}), (\ref{delta_FR_horizontal}), and (\ref{delta_FR_radial}), we get
\begin{align}
\label{gradient_super_adiabatic}
&C_p \overline{T} \left(\rho V_{r}\right)_0 \left(\overline{\nabla} - \nabla_a\right) \deriv{\ln \overline{p}}{r} 
= \left[\nu - (Q+1)\mu\right] \overline{\rho \epsilon} \left(\frac{\Delta T_0}{\overline{T}}\right)  \\
&- \frac{C_p \overline{T} }{\alpha l} \left(\frac{\Delta T_0}{\overline{T}}\right) \left(\rho V_{r}\right)_0 \frac{J_{1,M}}{J_{0,M}} -\alpha_2 \frac{k}{2} \frac{\left(\rho V_{r}^3\right)_0}{ l} \frac{J_{4,M}}{J_{0,M}} \nonumber \\
&-\frac{1}{r^2 } \deriv{}{r} \left\{ r^2 \, \chi_2 \deriv{\ln \left(\Delta T_0 / \overline{T}\right)}{r} \right\}
-\frac{1}{r^2 } \deriv{}{r} \left\{ r^2 \, \chi_2  \deriv{\ln \theta_M}{r}\right\} \frac{J_{2,M}}{J_{0,M}} \nonumber \\
&-\frac{1}{r^2 } \deriv{}{r} \left\{ r^2 \, \chi_1 \overline{F}_{R,r} \right\} + \chi_1 \overline{F}_{R,r} \deriv{\ln \theta_M}{r} \frac{J_{2,M}}{J_{0,M}} \nonumber \\
&+ \chi_2 \left\{\frac{J_{2,M}+J_{3,M}}{J_{0,M}}\left(\deriv{\ln \theta_M}{r}\right)^2 - \frac{J_{2,M}}{J_{0,M}}  \deriv{\ln \theta_M}{r} + \frac{2}{\pi} \frac{\sin \theta_M}{\theta_M r^2}  \frac{1}{J_{0,M}}\right\} \nonumber \, , 
\end{align}
where, for the entropy gradient, we used the relation 
\begin{align}
\deriv{\overline{S}}{r} = C_p (\overline{\nabla}-\nabla_a) \deriv{\ln \overline{p}}{r} \, .
\end{align}

If we consider \eq{gradient_super_adiabatic} in the overshooting zone above a convective core, because of the low radiative conductibility in stellar interior, the second and third terms of the right hand side will be the leading ones and we have with a very good approximation (this also holds deeper in the convective core)
\begin{align}
\label{gradient_super_adiabatic_overshoot}
&\left(\overline{\nabla} - \nabla_a\right)   
= \frac{H_p}{\alpha l} \left(\frac{\Delta T_0}{\overline{T}}\right) \frac{J_{1,M}}{J_{0,M}} + \alpha_2 \frac{k}{2} \frac{V_{r,0}^2 H_p}{C_p \overline{T} l} \frac{J_{4,M}}{J_{0,M}}  \, . 
\end{align}
In the overshooting layers $\Delta T_0 / \overline{T} <0$ and the first term is negative while the second one is of course positive. At the bottom of the overshooting region, we expect a slightly sub-adiabatic temperature gradient and therefore $\Delta T_0 / \overline{T} $ will become increasingly negative as the convective flux must then become increasingly negative there. On the other hand, the radial velocity $V_{r,0}$ can only decrease. Therefore, after remaining for a while close to the adiabatic value, the temperature gradient will become more and more sub-adiabatic. This will slow down the decrease of $\Delta T_0 / \overline{T} $ making it possible to have the temperature gradient varying progressively from the adiabatic value to the radiative one in a significant fraction of the overshooting zone. As pointed out by \cite{Zahn1991}, the neglected terms in \eq{gradient_super_adiabatic_overshoot} will be significant only in a thin layer on top of the overshooting zone.

It is also interesting to compare \eq{gradient_super_adiabatic} with its counterpart in the LMLT framework. There, $\epsilon=0$ and $\alpha_2=0$. The second term of the right hand side member is also present and the other ones correspond to $-\vec \nabla \cdot \Delta \vec F_R$ , which is approximated by 
\begin{align}
\label{simplified_66}
-\vec \nabla \cdot \Delta \vec F_R \simeq 
\frac{4ac}{3} \overline{\left(\frac{T^4}{\kappa \rho}\right)}\; \vec \nabla \cdot \vec \nabla \left(\frac{\Delta T_0}{\overline{T}}\right) \simeq 
\frac{4ac}{3 \mathcal{L}^2} \overline{\left(\frac{T^4}{\kappa \rho}\right)} \, .
\end{align} 
To recover Eq.~(42) in \cite{Henyey1965}, we must take $\mathcal{L} = \sqrt{2}\pi l_{\rm MLT}$. 
Therefore, we obtain
\begin{align}
V_r \left(\overline{\nabla}-\nabla_{\rm ad}\right)
\deriv{\ln \overline{p}}{r} = - \left[\frac{1}{\alpha \tau} + \omega_R \right] \left(\frac{\Delta T_0}{\overline{T}}\right)
\end{align}
with 
\begin{align}
\omega_R = \frac{4ac}{3} \overline{\frac{T^3}{C_p \kappa \rho \mathcal{L}^2}} \, .
\end{align}
The latter equation is identical to the usual one (except for our definition of the mean free path). The local solution of the set of equations obtained in this section simplified with the approximations $F_K=0$, $\theta_M$ constant, no advection, and also the approximate \eq{simplified_66}, allows us to recover the Böhm-Vitense cubic equation as obtained from the mean-field equations by \cite{Gabriel1974} or from the Lorenz solution or also the vortex model as shown by \cite{Arnett2011} and \cite{Smith2014}. 

\subsection{Advection velocity}

Finally, to close the system of equations, we need an additional equation to obtain the advection velocity or equivalently the radial mass flux $\left(\rho V_{r}\right)_0$. To this end, we start by considering \eq{eq_thermal_div_fluxes2}, and simply noting that 
\begin{align}
\Delta F_{K,r} + \Delta F_{C,r} = F_{K,r} + F_{C,r} - \left(\overline{F}_{K,r} + \overline{F}_{C,r}\right) \, , 
\end{align}
then, using \eq{mean_luminosity} and performing integration gives
\begin{align}
\label{advection_plume}
&J_{4,M} \deriv{}{r} \left[ r^2 \frac{k}{2} \left(\rho V_{r}^3\right)_0 \right]
+3 J_{5,M} r^2 \frac{k}{2} \left(\rho V_{r}^3\right)_0 \deriv{\ln \theta_M}{r} \nonumber \\
&+ \deriv{}{r}\left(r^2 F_{C,r,0}\right) J_{1,M} 
+ r^2 F_{C,r,0} \deriv{\ln \theta_M}{r} J_{6,M}   \\
&-(1-\cos \theta_M) \deriv{}{r} \left\{ r^2 \left[ C_p \overline{T} \left(\frac{\Delta T_0}{\overline{T}}\right) \left(\rho V_r\right)_0 G_1 
+ \frac{k}{2} \left(\rho V_r^3\right)_0 G_3 \right] \right\} \nonumber \\
&+ r \sin \theta_M \left( F_{K,\theta}\right)_{\theta_M} = r^2\frac{\rho C_p T}{\alpha l} \left(\frac{\Delta T_0}{\overline{T}}\right) J_{1,M}
+ \alpha_2 \frac{k}{2} \frac{\left(\rho V_{r}^3\right)_0}{l} r^2 J_{4,M} \, , \nonumber
\end{align}
where, using \eq{def_convective_flux}, we used the notation $F_{C,r,0} \equiv \overline{T} \Delta S_0 \left(\rho V_r\right)_0$.

\section{System to be integrated}
\label{section_5}

In this section, we summarize the equations to be integrated. Obviously, they have to be complemented by an equation of state, the  equations for the kinetic of nuclear reactions, and the three usual equations, namely
\begin{align}
\deriv{m}{r} &= 4\pi\rho r^2 \\
\deriv{p}{r} &= -G\frac{m}{r^2} \rho \\
\deriv{L}{r} &= 4 \pi \rho r^2 \left[\epsilon_N - T\deriv{S}{t}\right] = 4\pi \rho r^2 \epsilon \, .
\end{align}
In radiative layers, we also have to add   
\begin{align}
\deriv{T}{r} = - \frac{4ac}{3} \frac{\kappa \rho}{T^3} \frac{L(r)}{4\pi r^2} \, .
\end{align}

In the convective core we have first to take into account the equations integrated over a spherical surface. More precisely, they are;
\begin{itemize}
\item   the equation of continuity (Eq.~\ref{mass_flux_AD1} or Eq.~\ref{mass_flux_AD2})
\begin{align}
\label{continuity_sphere_final}
N J_{0,M} + A_D N_D  J_{0,D}  \quad {\rm or} \quad J_{0,M} + A_D  J^\prime_{0,D} = 0 \, , 
\end{align}
\item the equation of mechanical energy conservation (Eq.~\ref{mean_mechanical}) 
\begin{align}
\label{mechanical_sphere_final}
\frac{1}{r^2} \deriv{}{r} \left( \frac{k r^2 }{2} \left(\rho V_r^3\right)_0 G_3\right) &= 
-\left(\frac{\Delta \rho_0}{\overline{\rho}}\right) \left(\rho V_r\right)_0 g G_1 \nonumber \\
&- \frac{ k \left(\rho V_r^3\right)_0}{l} G_4 \, , 
\end{align}
\item  the equation of thermal energy conservation (Eq.~\ref{mean_luminosity}) 
\begin{align}
\label{luminosity_sphere_final}
\frac{\overline{L}(r)-\overline{L}_R(r)}{4\pi r^2} = C_p \Delta T_0 \left(\rho V_r\right)_0 G_1 
+ \frac{k}{2} \left(\rho V_r^3\right)_0 G_3 \, , 
\end{align}
with \eq{defs_Gi} for the definition of the coefficients $G_i$. 
\end{itemize}

The same equations integrated over the “horizontal” section of a plume, together with the one giving the advection velocity, are also to be taken into account. Namely;
\begin{itemize}
\item the equation of continuity (Eq.~\ref{continuity_plume})
\begin{align}
\label{continuity_plume_final}
&J_{0,M} \deriv{}{r} \left[r^2 \left(\rho V_r\right)_0 \right] 
+ J_{2,M} r^2 \left(\rho V_r\right)_0 \deriv{\ln \theta_M}{r} \nonumber \\
&+ r\sin \theta_M \left(\rho V_\theta\right)_{\theta_M} = 0 \, .
\end{align}
\item the equation of mechanical energy conservation (Eq.~\ref{mechanical_plume})
\begin{align}
\label{mechanical_plume}
&J_{4,M} \deriv{}{r} \left[r^2\frac{k}{2} \left(\rho V_r^3\right)_0\right] + \frac{3}{2} k r^2 \left(\rho V_{r}^3\right)_0 \deriv{\ln \theta_M}{r} J_{5,M}  \\
&+ r \left(\sin \theta \, F_{K,\theta}\right)_{\theta_M} 
= - r^2 \left[\frac{\Delta \rho_0}{\overline{\rho}} \left(\rho V_r\right)_0 g J_{1,M} 
+  \frac{k  \left(\rho V_r^3\right)_0}{l} J_{4,M}\right] \, , \nonumber
\end{align} 
where 
\begin{align}
\left(F_{K,\theta}\right)_{\theta_M} = \left(\rho V_\theta^3\right)_{\theta_M} \, .
\end{align}
\item the super adiabatic gradient (Eq.~\ref{gradient_super_adiabatic}) 
\begin{align}
\label{gradient_super_adiabatic_final}
&C_p \overline{T} \left(\rho V_{r}\right)_0 \left(\overline{\nabla} - \nabla_a\right) \deriv{\ln \overline{p}}{r} 
= \left[\nu - (Q+1)\mu\right] \overline{\rho \epsilon} \left(\frac{\Delta T_0}{\overline{T}}\right)  \\
&- \frac{C_p \overline{T} }{\alpha l} \left(\frac{\Delta T_0}{\overline{T}}\right) \left(\rho V_{r}\right)_0 \frac{J_{1,M}}{J_{0,M}} -\alpha_2 \frac{k}{2} \frac{\left(\rho V_{r}^3\right)_0}{ l} \frac{J_{4,M}}{J_{0,M}} \nonumber \\
&-\frac{1}{r^2 } \deriv{}{r} \left\{ r^2 \, \chi_2 \deriv{\ln \left(\Delta T_0 / \overline{T}\right)}{r} \right\}
-\frac{1}{r^2 } \deriv{}{r} \left\{ r^2 \, \chi_2  \deriv{\ln \theta_M}{r}\right\} \frac{J_{2,M}}{J_{0,M}} \nonumber \\
&-\frac{1}{r^2 } \deriv{}{r} \left\{ r^2 \, \chi_1 \overline{F}_{R,r} \right\} + \chi_1 \overline{F}_{R,r} \deriv{\ln \theta_M}{r} \frac{J_{2,M}}{J_{0,M}} \nonumber \\
&+ \chi_2 \left\{\frac{J_{2,M}+J_{3,M}}{J_{0,M}}\left(\deriv{\ln \theta_M}{r}\right)^2 - \frac{J_{2,M}}{J_{0,M}}  \deriv{\ln \theta_M}{r} + \frac{2}{\pi} \frac{\sin \theta_M}{\theta_M r^2}  \frac{1}{J_{0,M}}\right\} \nonumber \, , 
\end{align}
where  
\begin{align}
\deriv{\overline{S}}{r} = C_p (\overline{\nabla}-\nabla_a) \deriv{\ln \overline{p}}{r} \, .
\end{align}
\item and finally the advection velocity (Eq.~\ref{advection_plume})
\begin{align}
\label{advection_plume_final}
&J_{4,M} \deriv{}{r} \left[ r^2 \frac{k}{2} \left(\rho V_{r}^3\right)_0 \right]
+3 J_{5,M} r^2 \frac{k}{2} \left(\rho V_{r}^3\right)_0 \deriv{\ln \theta_M}{r} \nonumber \\
&+ \deriv{}{r}\left(r^2 F_{C,r,0}\right) J_{1,M} 
+ r^2 F_{C,r,0} \deriv{\ln \theta_M}{r} J_{6,M}   \\
&-(1-\cos \theta_M) \deriv{}{r} \left\{ r^2 \left[ C_p \overline{T} \left(\frac{\Delta T_0}{\overline{T}}\right) \left(\rho V_r\right)_0 G_1 
+ \frac{k}{2} \left(\rho V_r^3\right)_0 G_3 \right] \right\} \nonumber \\
&+ r \sin \theta_M \left(F_{K,\theta}\right)_{\theta_M} = r^2\frac{\rho C_p T}{\alpha l} \left(\frac{\Delta T_0}{\overline{T}}\right)  J_{1,M}
+ \alpha_2 \frac{k}{2} \frac{\left(\rho V_{r}^3\right)_0}{l} r^2 J_{4,M} \, .\nonumber
\end{align}
\end{itemize}
It is worth noticing that \eq{continuity_sphere_final} and \eq{luminosity_sphere_final} are algebraic and not differential. 

Consequently, we have a total of ten unknowns (i.e. ($m, p, T, L, (\rho V_r )_0, \Delta \rho_0 /\overline{\rho}, V_{\theta_M}, \theta_M, N$) and $A_D$ (we may exclude Eq.~\ref{continuity_sphere_final} of the number of equations and then not consider $A_D$ as a variable). Also, we could have added the extra variable $\theta_{M,D}$ given either by $N\left(1-\cos\theta_M\right)+N_D \left(1-\cos \theta_{MD}\right)=2$ or $N (1-\cos \theta_{M,D})=2$ (see Eq.~\ref{eq_A19}). Also, in the first case we can consider that $N_D \neq N$ and then $N_D$ will be a free parameter. 
However, we have as in the LMLT several parameters, namely the mixing length $l_{\rm MLT}$ and other more or less arbitrary constants. 
 Therefore, we have as many equations as unknowns and we do not need the usual Taylor’s closure hypothesis. Finally, those equations have to be complemented by the boundary conditions at the centre. 

\section{Remarks concerning $\alpha$ and $\alpha_2$}
\label{section_6}

Before discussing the boundary conditions, it is worth considering the treatment of $\alpha$ and $\alpha_2$. While most often considered as constant parameters and more precisely equal in up- and down-flows, we will show that this is more a convention rather than a physically-grounded assumption, and we thus propose an alternative way of considering them.

First, let us consider \eq{eq_thermal_div_fluxes2},
\begin{align}
\label{eq_thermal_div_fluxes2_tmp}
\vec \nabla \cdot \left( \Delta \vec F_K + \Delta \vec F_C \right) = \frac{\rho T \Delta S \left\vert V_r \right\vert }{\alpha l} + \alpha_2 \frac{k}{2} \frac{\rho V_r^3}{l} \; .
\end{align}
The average over a spherical surface of \eq{eq_thermal_div_fluxes2_tmp} being null, $\alpha$ and $\alpha_2$ must have different values in up- and down-flows. This implies 
\begin{align}
\label{alpha_*}
&\frac{C_p \rho T}{\alpha} \left(\frac{\Delta T_0}{\overline{T}}\right) V_{r,0} N J_{1,M}
+ \alpha_2 \frac{k}{2} \left(\rho V_r^3\right)_0 N J_{4,M}  \\
&= \frac{C_p \rho T \, l}{\alpha_D \, l_D} \left(\frac{\Delta T_0}{\overline{T}}\right) A_D^2 V_{r,0} N_D J_{1,D}
+ \alpha_{2,D} \left\vert A_D^3 \right\vert \frac{k_D \, l}{2\, l_D} \left(\rho V_r^3\right)_0 N_D J_{4,D} \, , \nonumber 
\end{align}
where, in the latter equation, $\alpha,\alpha_2$ and $\alpha_D,\alpha_{2,D}$ are associated with the up- and down-flows, respectively. 
We draw attention to the fact that we need the mean temperature gradients obtained from the up and down flows to be the same.

Except in very small regions close to each boundary of the convective core but much thinner than one layer there, because of the very low thermal conductivity deep inside stars and also because $\epsilon \ll C_p \overline{T}$, \eq{gradient_super_adiabatic_final} reduces with a very good accuracy to
\begin{align}
&V_{r,0} \left(\overline{\nabla} - \nabla_{a}\right) \deriv{\ln \overline{p}}{r} = \nonumber \\
& - \frac{1}{\alpha l} \left(\frac{\Delta T_0}{\overline{T}}\right) \left\vert V_{r,0}\right\vert \frac{J_{1,M}}{J_{0,M}} 
- \alpha_2 \frac{k}{2 l} \frac{\left(\rho V_r^3\right)_0}{C_p \overline{\rho} \overline{T}} \frac{J_{4,M}}{J_{0,M}}  \, .
\end{align}
Therefore we must have for the down-flow 
\begin{align}
&V_{r,0} \left(\overline{\nabla} - \nabla_{a}\right) \deriv{\ln \overline{p}}{r} = \nonumber \\
& - \frac{\left\vert A_D \right\vert}{\alpha_D l_D} \left(\frac{\Delta T_0}{\overline{T}}\right) \left\vert V_{r,0}\right\vert \frac{J_{1,D}}{J_{0,D}} 
- A_D^2 \alpha_{2,D} \frac{k_D}{2 l_D} \frac{\left(\rho V_r^3\right)_0}{C_p \overline{\rho} \overline{T}} \frac{J_{4,D}}{J_{0,D}}  
\end{align}
thus  
\begin{align}
\label{alpha_**}
&- \frac{1}{\alpha l} \left(\frac{\Delta T_0}{\overline{T}}\right) \left\vert V_{r,0}\right\vert \frac{J_{1,M}}{J_{0,M}} 
- \alpha_2 \frac{k}{2 \, l} \frac{\left(\rho V_r^3\right)_0}{C_p \overline{\rho} \overline{T}} \frac{J_{4,M}}{J_{0,M}}  
= \nonumber \\
&- \frac{\left\vert A_D \right\vert}{\alpha_D l_D} \left(\frac{\Delta T_0}{\overline{T}}\right) \left\vert V_{r,0}\right\vert \frac{J_{1,D}}{J_{0,D}} 
- A_D^2 \alpha_{2,D} \frac{k_D}{2\,l_D} \frac{\left(\rho V_r^3\right)_0}{C_p \overline{\rho} \overline{T}} \frac{J_{4,D}}{J_{0,D}}  \, .
\end{align}
Equations (\ref{alpha_*}) and (\ref{alpha_**}) form a system connecting the two sets of variable (i.e. $\alpha,\alpha_2$ and $\alpha_D,\alpha_{2,D}$). Obviously, if one decides that $\alpha$ and $\alpha_2$ are constants then the other pair must be variable. However, there is no reason to treat $\alpha,\alpha_2$ as constants and $\alpha_D,\alpha_{2,D}$ as variables (or the other way around). The only way to avoid this is if these variables are the solution of the following two systems:
\begin{align}
&\frac{C_p \overline{\rho} \overline{T}}{\alpha l} \left(\frac{\Delta T_0}{\overline{T}}\right)
N J_{1,M} + \alpha_2 \frac{k}{2} \frac{\overline{\rho} V_{r,0}^2}{l} N J_{4,M} = C_1(r) \\
& \frac{1}{\alpha l}  \left(\frac{\Delta T_0}{\overline{T}}\right) \frac{J_{1,M}}{J_{0,M}} 
+ \alpha_2 \frac{k}{2} \frac{\overline{\rho} V_{r,0}^2}{C_p \overline{\rho} \overline{T} l} \frac{J_{4,M}}{J_{0,M}} = C_2(r)
\end{align}
and 
\begin{align}
&\frac{C_p \overline{\rho} \overline{T}}{\alpha_D l_D} \left(\frac{\Delta T_0}{\overline{T}}\right) N_D A_D^2 J_{1,D} + \alpha_{2,D} \frac{k_D}{2} \frac{\overline{\rho} V_{r,0}^2}{l_D} \left\vert A_D^3 \right\vert N_D J_{4,D} = C_1(r) \\
& \frac{1}{\alpha_D l_D}  \left(\frac{\Delta T_0}{\overline{T}}\right) 
\left\vert A_D \right\vert \frac{J_{1,D}}{J_{0,D}} 
+ \alpha_{2,D} \frac{k_D}{2} \frac{\overline{\rho} V_{r,0}^2 A_D^2}{C_p \overline{\rho} \overline{T} l_D} \frac{J_{4,D}}{J_{0,D}} = C_2(r),
\end{align}
where $C_1(r)$ and $C_2(r)$ are the new arbitrary coefficients, constant over a spherical surface.

The coefficients $C_1(r)$ and $C_2(r)$ still depend on the radius even within the LMLT framework. This can be fixed by introducing to other coefficients, namely
\begin{align}
C_1(r) &= \frac{C_p \overline{\rho T}}{l} \left(\frac{\Delta T_0}{\overline{T}}\right) C_1^\prime(r) \\
C_2(r) &= \frac{1}{l} \left(\frac{\Delta T_0}{\overline{T}}\right) C_2^\prime(r) \, . 
\end{align} 
With this choice, within the LMLT, $C_1^\prime$ and $C_2^\prime$  no longer depend on the radius and are constants. Indeed, within the LMLT, one has
\begin{align}
 \left(\frac{\Delta T_0}{\overline{T}}\right) \propto \frac{V_{r,0}^2}{C_p \overline{T} \nabla_{a}} \, ,
\end{align}
since the mixing length is proportional to
the pressure scale height. Moreover, one has $A_D=-1$, $N J_{1,M} = N_D J_{1,D}$, $J_{4,M} = N_D J_{4,D}$, $J_{1,M}/J_{0,M} = J_{1,D}/J_{0,D}$, $J_{4,M}/J_{0,M} = J_{4,D}/J_{0,D}$, the latter being constants. 
We see that the two sets of equations are identical and the two sets of unknowns are equal. However, if $\alpha$ and $\alpha_2$ are constants, then $\alpha_D$ and $\alpha_{2D}$ will vary only
slightly as $\nabla_a$. This justifies once more the absence of the term in $\alpha_2$ in the LMLT. 
There is a slight difference with the usual LMLT, which results from a different choice of the “horizontal” variation of the variables. In the LMLT, the function $h\left(\theta,\phi\right)$ is assumed to be constant in the up- ($h\left(\theta,\phi\right)=1$) and down-flows ($h\left(\theta,\phi\right)=-1$). This finally suggests that we should take $C_1^\prime$ and $C_2^\prime$ as true constants independent of the distance to the centre.

\section{Boundary conditions at the centre}
\label{section_7}

Boundary conditions at the centre of convective cores constitute a non-trivial issue even if it is a coordinate singularity rather than a physical one. Regarding 3D hydrodynamical simulations, it is difficult to obtain the temperature gradient since the 3D hydrodynamical simulations are not yet able to reach the thermal relaxation as its characteristic time is very long. Also, for the vertical component of the velocity, two recent numerical simulations by \cite{Gilet2012} and \cite{Cai2011} adopt different boundary condition. The first one considers $V_r(0)=0$ while the second one considers a non-vanishing value. 

In this section, we provide boundary conditions for our model but we would like to stress that the discussion that follows is closely linked to our hypothesis that the plumes theory we decided to develop is valid down to the centre. One of the consequences of this hypothesis is that the equations have a singularity there and that the behaviour of the variables of the problem close to it is given by the requirement that the solution is regular. Of course our hypothesis can be put in question as the physics of the problem does not necessarily require a singularity at the centre. One could for instance prefer the kind of behaviour adopted by \cite{Cai2011}. Nevertheless we may reasonably hope that this choice will not significantly modify the solution in the outer part of the convective core where the temperature gradient departs significantly from the adiabatic value. The same point of view justifies our decision to ignore the very small super adiabatic central core we will find in the discussion here below and to assume that the central regions of the core are adiabatic.

\subsection{The temperature gradient at the centre}

Let us now consider \eq{gradient_super_adiabatic} (or Eq.~\ref{gradient_super_adiabatic_final}). It contains several terms that may lead to a singularity at the centre. The second to last of the right hand side member is the most singular one, so let us first compare it to the fourth one as these terms are the two components of the divergence of the radiative flux convective fluctuation. Clearly, close to the centre the non-radial component will dominate, because the “horizontal” extent of a plume becomes very small, so that the “horizontal” temperature gradient is very large. This term is still  larger than the radial one far from the centre. Indeed, we can roughly write the radial term such that
\begin{align}
-\frac{1}{C_p \overline{T} \overline{\rho}}\frac{1}{r^2 } \deriv{}{r} \left\{ r^2 \, \chi_2 \deriv{\ln \left(\Delta T_0 / \overline{T}\right)}{r}  \right\}\simeq\frac{4ac}{3} \frac{\overline{T}^3}{C_p \kappa \overline{\rho}^2 H_p^2} \left(\frac{\Delta T_0}{\overline{T}}\right) \, .
\end{align}
Therefore, the radial term is the leading one if
\begin{align}
\frac{H_p}{r} < \sqrt{\frac{\pi}{2}  \frac{\theta_M}{\sin \theta_M} J_{0,M}} \; \;  {\rm or} \; \;
\left\vert \deriv{\ln p}{\ln r} \right\vert > \left(\frac{\pi}{2}  \frac{\theta_M}{\sin \theta_M} J_{0,M}\right)^{-1/2} \, .
\end{align}
Consequently, except for large values of $\theta_M$ approaching one, this condition will not be fulfilled in most of the radiative cores so that we have to take the non-radial component into account. However, because of the low conductivity of the gas deep in stars, the second term of the right hand side member in \eq{gradient_super_adiabatic}  (or Eq.~\ref{gradient_super_adiabatic_final}) is expected to become quickly the leading one.

Finally, we conclude that, because the radiative dissipation through the “horizontal” component will be important in the vicinity of the centre, our formalism leads to a small non-adiabatic core near the centre. At first, we will estimate the extent of the non-adiabatic convective core.

Close to the centre, \eq{gradient_super_adiabatic} reduces to
\begin{align}
\label{gradient_cond_centre_tmp}
 V_{r,0} \left(\overline{\nabla} - \nabla_a\right) \deriv{\ln \overline{p}}{r} 
=  - \frac{8 ac}{3 \pi} \frac{\sin \theta_M}{\theta_M r^2}  \overline{\left(\frac{T^3}{C_p \kappa \rho^2}\right)} \left(\frac{\Delta T_0}{\overline{T}}\right) \frac{1}{J_{0,M}} \, , 
\end{align}
which must be solved together with \eq{mechanical_sphere_final} and \eq{luminosity_sphere_final} where 
\begin{align}
\label{defs_fluxes_tmp1}
\overline{F_{R,r}} &= \frac{\overline{L_R}}{4\pi r^2} = -\frac{4ac}{3} \overline{\left(\frac{T^4}{\kappa \rho}\right)} \; \overline{\nabla} \,  \deriv{\ln \overline{p}}{r} \\ 
\label{defs_fluxes_tmp2}
\overline{F_{r}} &= \frac{\overline{L}}{4\pi r^2} = -\frac{4ac}{3} \overline{\left(\frac{T^4}{\kappa \rho}\right)} \; \overline{\nabla}_R \,  \deriv{\ln \overline{p}}{r} \, ,
\end{align}
where $\overline{\nabla_R}$ is the radiative gradient. 

Thus, using Eqs.~(\ref{defs_fluxes_tmp1}) and (\ref{defs_fluxes_tmp2}) into \eq{gradient_cond_centre_tmp},  \eq{luminosity_sphere_final} becomes 
\begin{align}
&-\frac{4ac}{3} \overline{\left(\frac{T^4}{\kappa \rho}\right)} \deriv{\ln \overline{p}}{r} \left[\overline{\nabla}_R - \nabla_a - \left(\overline{\nabla} - \nabla_a\right)\right] 
= -\frac{4ac}{3} \overline{\left(\frac{T^4}{\kappa \rho}\right)}  \nonumber \\
&\times \left[ \left(\overline{\nabla}_R - \nabla_a\right) \deriv{\ln \overline{p}}{r} +\frac{8 ac}{3 \pi} \frac{\sin \theta_M}{\theta_M r^2}  \overline{\left(\frac{T^3}{C_p \kappa \rho^2}\right)} \left(\frac{\Delta T_0}{\overline{T}}\right) \frac{1}{J_{0,M} V_{r,0}} \right] \nonumber \\
&= C_p \overline{T} \left(\frac{\Delta T_0}{\overline{T}}\right) \left(\rho V_r\right)_0 G_1 
+ \frac{k}{2} \left(\rho V_r^3\right)_0 G_3.
\end{align}

To go further, it is necessary to take the Taylor expansion of $V_{r,0}$ and $\Delta T_0 / \overline{T}$ at least to the second order to ensure that the temperature gradient does not remain equal to the radiative one and becomes adiabatic at some point off-centre. To that end, we assume that 
\begin{align}
V_{r,0} &= C_V r^m \left[1- \left(\frac{r}{r_1}\right)^{2p}\right] \\
\left(\frac{\Delta T_0}{\overline{T}}\right) &= C_T r^n \left[1- \left(\frac{r}{r_2}\right)^{2q}\right] \, , 
\end{align}
with $n-m=3$, as demanded by \eq{gradient_cond_centre_tmp}. 

Further assuming that $G_1$ and $G_3$ are constant (which will be found justified later on) and that $l=\beta_1 r$ (which is justified by the large radiative dissipations), we get the relations
\begin{align}
&\frac{8 ac}{3 \pi} \frac{\sin \theta_M}{\theta_M r^2}  \overline{\left(\frac{T^3}{C_p \kappa \rho^2}\right)} \left(\frac{\Delta T_0}{\overline{T}}\right) \frac{1}{J_{0,M}} C_T 
= \frac{\overline{\nabla}_R - \nabla_a}{r H_p} J_{0,M} C_V \\
&\left[\left(3q+2\right)\frac{k}{2} G_3 + \frac{2}{\beta_1} G_4\right] C_V 
= \frac{\left(\overline{\nabla}_R - \nabla_a\right) Qg}{r} \frac{J_{0,M} G_1}{r H_p} \nonumber \\
&\times 
\left[\frac{8 ac}{3 \pi} \frac{\sin \theta_M}{\theta_M r^2}  \overline{\left(\frac{T^3}{C_p \kappa \rho^2}\right)} \left(\frac{\Delta T_0}{\overline{T}}\right)\right]^{-1} \\
&\left(\frac{r_1}{r_2}\right)^{12} = 2 \, , 
\end{align}
and 
\begin{align}
\label{eq_g1_cv2}
G_1 C_V^2 = \frac{2}{\pi} \frac{\sin \theta_M}{\theta_M} \left[ \frac{4ac}{3} 
\overline{\frac{T^3}{C_p \kappa \rho^2}}\right]^2 \frac{1}{J_{0,M}} \left(\frac{1}{r_1}\right)^{12}
\end{align}
together with the values $m=5, n=8$, and $p=q=6$.

We have computed these quantities for a few main-sequence models with an initial composition $X=0.72$ and $Z=0.015$ either close to the ZAMS or close to the middle main sequence (MMS). 
The values of $C_T$ and $C_V$ depend on the assumed values for the $G_i$ (see Table~\ref{table:1}). They allow nevertheless a good accuracy on $r_1$ because it appears at the power 12 in \eq{eq_g1_cv2}. The results are given in Table \ref{table:1}.

\begin{table*}
        \caption{First line gives the central hydrogen abundance of the model, the fifth gives the radius of the first off-centre point, the sixth the temperature difference between the centre and point 2. The meaning of the other symbols has already been defined in the text.}             
        \label{table:1}      
        \centering          
        \begin{tabular}{c c c c l l l }     
                \hline\hline       
                 & $1.3 M_\odot$ ZAMS & $1.3 M_\odot$ MMS & $1.5 M_\odot$ ZAMS & $1.5 M_\odot$ MMS & $10 M_\odot$ ZAMS& $10 M_\odot$ MMS\\ 
                \hline                    
                $X_C$ & 0.71890 & 0.35938  &   0.71900 & 0.35805 & 0.71903 &  0.35744  \\  
                $C_V$ & 2 $\times$10$^{-32}$ & 6.9 $\times$10$^{-32}$    & 8.76$\times$10$^{-34}$ & 5.1$\times$10$^{-35}$& 1.06$\times$10$^{-36}$ & 6.72$\times$10$^{-37}$\\
                $C_T$ & 2 $\times$10$^{-58}$ & 1.2 $\times$10$^{-57}$ & 3.54 $\times$10$^{-61}$ & 3.6 $\times$10$^{-61}$ & 3 $\times$10$^{-67}$ & 1.6 $\times$10$^{-68}$ \\
                $r_1 /R$ & 2.6 $\times$10$^{-5}$ & 1.86 $\times$10$^{-5}$ & 3.7 $\times$10$^{-5}$& 6.1 $\times$10$^{-5}$& 7.7 $\times$10$^{-5}$ & 7.7 $\times$10$^{-5}$ \\
                $r(2)/R$ & 0.0043687 & 0.0031341 & 0.0042997 & 0.0032806 & 0.0048923 & 0.0032806\\
                $T(1) -T(2)$ & 4 $\times$10$^{3}$ & 8 $\times$10$^{3}$ & 6 $\times$10$^{3}$ &  10$^{4}$ & 6 $\times$10$^{3}$ & 10$^{4}$ \\
                $\overline{\nabla}_R / \overline{\nabla}_a$ & 1.7809 & 1.6349 & 1.3325 &  2.7447 & 2.9679 & 2.7447 \\
                \hline                  
        \end{tabular}
\end{table*}

Though it is not very accurate to compute the extent of the super-adiabatic convective core with just a second-order expansion, we may expect that $r_1$ gives its right order of magnitude. Since   $r_1 \ll r(2)$ ($r(2)$ being the radius of the first off-centre point of the models) in all cases considered in Table.~\ref{table:1} and since $T(1)-T(2) \propto r (2)^2$ ($T(1)-T(2)$ being the temperature difference between the centre and the second mesh point of the models) is small, it will in practice make no difference (it will be smaller than the accuracy of the models computation) whether the temperature gradient is adiabatic down to the centre or whether there is a small super-adiabatic core. Moreover, it would be necessary to introduce many mesh points in that small domain in order to compute accurately the structure of this non-adiabatic core and particularly to follow the variation of the temperature gradient. Therefore we will ignore the latter and we will assume that $\overline{\nabla}_R = \nabla_a$ until we are far enough from the centre. Then, \eq{gradient_super_adiabatic_final} may be used to compute the departure from the adiabatic gradient. 
Nevertheless, this small super-adiabatic core as predicted by the plume theory as well as the nuclear reactions might help the formation of up-rising plumes.

\subsection{The boundary conditions}

We will now consider the Taylor expansion of the equations \eq{mechanical_sphere_final} to \eq{advection_plume_final}, assuming that the temperature gradient is adiabatic at the centre justified in the previous section.

Let us first recast \eq{mechanical_sphere_final} and \eq{luminosity_sphere_final} as
\begin{align}
\label{equation_y}
\deriv{y}{r} + \left[\frac{2 G_4}{l G_3} + \frac{\nabla_a}{H_p} \right] y = \frac{\nabla_a}{H_p} \frac{\overline{L}(r)-\overline{L}_R(r)}{4 \pi} \, ,
\end{align} 
with 
\begin{align}
y = \frac{k}{2} r^2 \left(\rho V_r^3\right)_0 G_3 \, .
\end{align} 

The second term in the bracket is negligible close to the centre and we will take again $l=\beta_1 r$. More generally, we suggest taking  $l=min[\beta_1 r,\beta_2 R_{CC},\beta_3 H_p],$ where $R_{CC}$ is the radius of the convective core and the $\beta_i$ are undetermined constants. 
This would be the usual way to define $l$ but, following the recent findings by \cite{Arnett2009,Arnett2015}, one can instead suggest to consider $l = min[\beta_1 r; \beta_2 R_{CC}]$. 
The solution of \eq{equation_y} is given by
\begin{align}
\label{cond_y}
y = r^2 \frac{k}{2} \left(\rho V_r^3\right)_0 G_3 = \frac{r \nabla_a}{H_p} \frac{\overline{L}(r)-\overline{L}_R(r)}{4 \pi \left[5+\frac{2G_4}{\beta_1 G_3}\right]}\, , 
\end{align}
which shows that $\left(\rho V_r\right)_0 \propto r$. As a result, the kinematic convective energy flux is negligible compared to the convective flux and $\left(\Delta T_0 / \overline{T}\right)$ is given by
\begin{align}
\label{cond_L}
\frac{\overline{L}(r)-\overline{L}_R(r)}{4 \pi r^2} &= - \frac{4ac}{3} \overline{\left(\frac{T^4}{\kappa \rho}\right)} \deriv{\ln \overline{p}}{r} \left[\overline{\nabla}_R - \nabla_a\right] \nonumber \\
&= C_p \overline{T} \left(\frac{\Delta T_0}{\overline{T}} \right) \left(\rho V_r\right)_0 G_1 \, , 
 \end{align}
and it is different from zero at the centre. Not surprisingly, we recover the same results as \cite{LoSchatzman1997}. 
 
To go further,  we consider a Taylor expansion of $V_{r,0}$ and $\Delta T_0 / \overline{T}$ and we will keep the notations as in the previous section,
\begin{align}
 V_{r,0}=C_V r^m \\
 \left(\frac{\Delta T_0}{\overline{T}} \right) =C_T r^n, 
\end{align}
but now m=1 and n=0 as imposed by \eq{cond_y} and \eq{cond_L}.
 
 We now consider the equations for a plume. Let us first consider \eq{continuity_plume}. The second term of the equation is negligible as it is at least proportional to $r^3$ so that we obtain 
 \begin{align}
 \left(m+2\right) J_{0,M} \left(\rho V_r\right)_0 + \sin \theta_M \left(\rho V_\theta\right)_{\theta_M} = 0 
 \end{align}
 or 
 \begin{align}
 \left(\rho V_\theta^3\right)_{\theta_M} = - \left(\frac{\left(m+2\right) J_{0,M}}{\sin \theta_M}\right)^3 \left(\rho V_r^3\right)_0.
 \end{align}
 We see that close to the centre the advection is directed towards the interior of the plume.
 
 Let us now turn to \eq{mechanical_plume}. It leads to
 \begin{align}
&3 r J_{5,M} \left[\frac{k}{2} \deriv{\ln \theta_M}{r}\right] = 
\sin \theta_M \left(\frac{\left(m+2\right) J_{0,M}}{\sin \theta_M}\right)^3  
- \left(3m+2\right) J_{4,M} \frac{k}{2}  \nonumber \\
&+ J_{1,M} Q \frac{\overline{p}}{\overline{\rho} r H_p} \frac{C_T}{C_V^2} - \frac{1}{\beta_1} J_{4,M} k \, , 
 \end{align}
 and to avoid the singularity of ${\rm d}\ln\theta_M/{\rm d}r,$ we must have
 \begin{align}
 &\sin \theta_M \left(\frac{\left(m+2\right) J_{0,M}}{\sin \theta_M}\right)^3 
 - \left(3m+2\right) J_{4,M} \frac{k}{2} + J_{1,M} Q \frac{\overline{p}}{\overline{\rho} r H_p} \frac{C_T}{C_V^2} \nonumber \\
 &- \frac{1}{\beta_1} J_{4,M} k = 0 \, .
 \end{align}
 This equation shows that $\theta_M$ is constant over the domain of the Taylor expansion. Also through the $G_i$ appearing in the definition of $C_T$ and $C_V$, the three variables $N, N_D,$ and $\theta_{MD}$ appear in this equation. There is a relation between $N, N_D, \theta_M,$ and $\theta_{MD}$ as we have either $N\left(1-\cos \theta_M\right) +N_D \left(1-\cos \theta_{MD}\right)=2$ or  $N(1-\cos \theta_{MD} )=2$ (see Eq.~\ref{eq_A19}). 
 
 Finally, we consider \eq{advection_plume_final}. The leading terms of this equation include $F_{C,r,0}$ , which gives 
 \begin{align}
 J_{6,M} r \deriv{\ln \theta_M}{r} &=- \left(m+n+2\right) J_{1,M} \nonumber \\
 &+ \left(m+n+2\right) \left(1- \cos \theta_{M}\right)G_1 + \frac{J_{1,M}}{\alpha \beta_1} \, . 
 \end{align}
Proceeding as previously, one has 
 \begin{align} 
\left(m+n+2\right) \left(1- \cos \theta_{M}\right)G_1 = -\frac{J_{1,M}}{\alpha \beta_1} + \left(m+n+2\right) J_{1,M} \, .
 \end{align}
 This equation provides the second relation between $\theta_M$ and $N$. We see that the values of $N$ and $\theta_M$ depend on two poorly known parameters $\alpha$ and $\beta_1$. 
 
 It could also be tempting to cancel the terms in $\left(\rho V_{r}^3\right)_0$ so that 
  \begin{align}
 J_{5,M} r \deriv{\ln \theta_M}{r} &=-\left(3m+2\right) J_{4,M} + \left(3m+2\right) \left(1-\cos \theta_M\right) G_3 \nonumber \\
 &+ \frac{2}{k} \sin \theta_M \left(\frac{\left(q+2\right) J_{0,M}}{\sin \theta_M}\right)^3 + \frac{\alpha_2}{\beta_1} J_{4,M} \, .
   \end{align}
Thus, 
\begin{align}
\label{boundary_last}
&-\left(3m+2\right) J_{4,M} + \left(3m+2\right) \left(1-\cos \theta_M\right) G_3 + \frac{\alpha_2}{\beta_1} J_{4,M} \nonumber \\
&+ \frac{2}{k} \sin \theta_M \left(\frac{\left(m+2\right) J_{0,M}}{\sin \theta_M}\right)^3  = 0 \, .
\end{align}
This equation would enable the computation of $\alpha_2$. However, we have to remember that the solutions for $V_{r,0}$ and therefore for $\Delta T_0 / \overline{T}$ are valid to the first order only and that if a solution taking higher-order terms had been computed, the lower order terms in $\left(\rho V_{r}^3\right)_0$  would be mixed with higher order terms in $F_{C,r,0}$ so that \eq{boundary_last} is not valid and $\alpha_2$ remains a free parameter.

\section{Case of an expanding core with a discontinuity of chemical composition at its surface}
\label{section_8}

During some fraction of the main-sequence phase in stars slightly more massive than the Sun and in many stars during central helium burning, the expanding core creates a discontinuity of chemical composition at its surface. In that situation, rising convective matter will reach the discontinuity with a finite radial velocity and will try to overshoot above it, but the difference between the densities on both sides of the discontinuity will soon become very large compared to the convective density fluctuations. That is to say, if $\rho_+$  and $\rho_-$ are respectively the densities above and below the discontinuity, one has $\left\vert \rho_+ - \rho_- \right\vert \gg \left\vert \Delta \rho_{0,-} \right\vert$. Therefore, if we consider the radial motion of matter above the discontinuity, it is given with a good approximation by 
\begin{align}
\frac{1}{r^2} \deriv{}{r} \left( r^2 \rho_- V_{r,0}^2 \right) = (\rho_+ - \rho_-) \frac{G m(r)}{r^2} \, , 
\end{align}
as it is essentially slowed down by the buoyancy force, which is much larger that the dissipation term. This gives a penetration length, $\Delta r$, given by 
\begin{align}
\frac{V_{r,0,-}^2}{c_-^2} = -\frac{1}{\Gamma_1} \frac{(\rho_+ - \rho_-)}{\rho_-} \frac{\Delta r}{H_{p-}} \, , 
\end{align}
where $V_{r,0,-}$ and $c_-$ are respectively the radial velocity and the sound speed just below the discontinuity. As $V_{r,0,-}^2 / c_-^2 \simeq 10^{-6}-10^{-7}$ using orders of magnitude given by the LMLT and $(\rho_+ - \rho_-) / \rho_- \simeq 10^{-1}-10^{-2}$, $\Delta r / H_{p-} \simeq 10^{-4}$ and the overshooting distance will be very small. As a matter of fact, rather than to overshoot, the rising material will crush against the discontinuity as the ratio $(\rho_+ - \rho_-) / \Delta \rho_{0,-}$ is enormous. This will generate internal gravity waves  \citep[see][]{Meakin2007a,Meakin2007b,Arnett2009} propagating along the discontinuity and producing some mixing, which will be proportional to $(X_+-X_- )$ for main-sequence stars or to $(Y_+ -Y_-)$ during central helium burning and to $[S_{M} (\rho V_r )_M ]_-$  where $S_{M}$ is the fraction of the spherical surface coinciding with the discontinuity occupied by rising material and $(\rho V_r)_M$ is the average radial component of the impulsion of rising material at the position of the discontinuity of chemical composition.

The equations obtained in the preceding sections are still valid below this complex mixing layer coinciding just with the discontinuity, and therefore provide the value of $\Delta \rho_{0,-}$ and $V_{r,0,-}$. However, one must also determine the evolution of the mass of the convective core. To do so, we will consider stars during the main-sequence  phase, but the transposition to the central helium burning phase is straightforward. 

The variation of the mass of the convective core can be written as 
\begin{align}
\label{def_mnc}
\deriv{M_{\rm NC}}{t} = 4\pi r^2 \frac{F_{X,r}}{\left(X_+ - X_-\right)} \, , 
\end{align}
where $F_{X,r}$ is the radial component of the hydrogen flux, which is given by 
\begin{align}
\label{def_flux}
F_{X,r} &= - \frac{\Delta M}{M} \left( X_+ - X_- \right) \left[ S_M \left(\rho V_r\right)_M \right]_- \nonumber \\
&= X_- \left[ S_M \left(\rho V_r\right)_M \right]_- + X \left[ S_D \left(\rho V_r\right)_D \right]_- \nonumber \\
&= (X_- - X) \left[ S_M \left(\rho V_r\right)_M \right]_- \, ,
\end{align}
where the subscript $D$ corresponds to down-moving matter and $X$ is the abundance in descending material after having interacted and exchanged some matter with the material above the discontinuity and it reads 
\begin{align}
\label{eq_discont_X}
X = X_- + \frac{\Delta M}{M} \left(X_+ - X_-\right) \, , 
\end{align}
where $\Delta M / M$ is the constant of proportionality. 
We recall that $\left(\rho V_r\right)_M$ and $\left(\rho V_r\right)_D$ stand for averaged values and therefore we have 
\begin{align}
\label{recall_mass_flux}
\left(\rho  V_r\right)_M &= \left(\rho  V_r \right)_0 \frac{2}{\theta_M^2} \int_{0}^{\theta_M} \cos\left(\frac{\pi}{2}\frac{\theta}{\theta_M}\right) \sin \theta \, {\rm d}\theta \nonumber \\ 
&=\frac{2J_{0,M}}{\theta_M^2} \left(\rho V_r \right)_0 = \frac{2J_{0,M}}{\theta_M^2} \left(\rho V_r \right)_{0,-} \, .
\end{align}

Finally, using \eq{def_mnc} together with \eq{def_flux} and \eq{recall_mass_flux}, one gets 
\begin{align}
\label{final_evolve_mass}
\deriv{M_{\rm NC}}{t} = 4\pi r^2 \frac{F_{X,r}}{\left(X_+ - X_-\right)} 
= 4\pi r^2 \frac{\Delta M}{M} \frac{2J_{0,M}}{\theta_M^2} S_{M,-}\left(\rho \vec V\right)_{0,-} \, .
\end{align}

From \eq{final_evolve_mass}, we can obtain the value of  $\Delta M / M$, which allows a growth rate of the core mass 
given by the location of the chemical composition discontinuity such that, at its lower side, that is to say in the convective core, $L_R=L$ is verified. 
The value obtained is the value assumed implicitly in stellar evolution. It is also a minimal one (even though one could also consider evolutions with undershooting). This is what we consider to be an evolution without overshooting. If a larger value is chosen, it will lead to an evolution with overshooting and thus models with larger convective cores and also $L_R>L$ at the location of the chemical discontinuity. However, $\left(\rho  V_r \right)_{0,-}$ decreases with increasing overshooting and this could quickly require unreasonable values of $\Delta M / M$ to have the convective core mass still growing fast enough. This leads naturally to a maximum overshooting. We see that the mechanism that determines the extent of the overshooting is different in shrinking and in expanding convective cores. In shrinking cores, the extent of the overshooting layers is given by the rate of slowing down of the rising material computed in the frame of a coherent theory. In an expanding core, it is given by the ability of the rising material to erode the chemical discontinuity; the rate of this process fixes the extent of the overshooting region and there is no argument to choose it as proportional to the pressure scale height.

We notice from \eq{mean_luminosity} that if the kinetic energy flux is positive ($G_3>0$) at the location of the chemical composition discontinuity where $L \le L_R$, the convective flux must be negative and  $\Delta \rho_M /\overline{\rho}>0$ there. If we take \cite{Spruit2015}'s suggestion that the density fluctuation of the descending material must be positive, we can get a maximum value of  $\Delta M / M$, assuming that $\Delta \rho_D / \overline{\rho}=0$ since, if it was negative, the material would not move down spontaneously.

Since $\overline{p} = p_D$ , if we expand this equation to the first order taking into account that the chemical composition of the down-moving material is different from that of the convective core, assuming that the rising material has the same composition as the average medium, we get 
\begin{align}
\frac{\overline{\rho}-\rho_D}{\overline{\rho}} = -Q \frac{\overline{T}-T_D}{\overline{T}} 
- \frac{\left(\derivp{\ln p}{X}\right)_{T,\rho}-\left(\derivp{\ln p}{Y}\right)_{T,\rho}}{\left(\derivp{\ln p}{\ln \rho}\right)_{T,X_i}} 
\left(X_M-X_D \right) =0 \, , 
\end{align}
and
\begin{align}
Q \frac{\overline{T}-T_D}{\overline{T}} &= 
\frac{\left(\derivp{\ln p}{X}\right)_{T,\rho}-\left(\derivp{\ln p}{Y}\right)_{T,\rho}}{\left(\derivp{\ln p}{\ln \rho}\right)_{T,X_i}} 
\left(X_D-X_M \right) \nonumber \\
&= \frac{\left(\frac{\ln p}{\ln \mu^{-1}}\right)_{T,\rho} \mu \left[ 2-3/4 \right]}{\left(\derivp{\ln p}{\ln \rho}\right)_{T,X_i}} \left(X_D-X_M\right) \, .
\end{align}
For a perfect gas, 
\begin{align}
\label{eq_discont_XD}
(X_D-X_M) = \frac{4}{5\mu} \frac{\overline{T}-T_D}{\overline{T}} \, .
\end{align}
This requires a hypothesis concerning $\frac{\overline{T}-T_D}{\overline{T}}$ and it seems reasonable to assume that
\begin{align}
\label{eq_discont_T}
\frac{\overline{T}-T_D}{\overline{T}} = - \frac{\overline{T}-T_M}{\overline{T}} >0 \, .
\end{align}
If we notice that $X$ and $X_-$ in \eq{eq_discont_X} correspond to $X_D$ and $X_M$ in \eq{eq_discont_XD} and since $(X_+ - X_-)$ is given by the model, these two equations provide us with a value of $\Delta M/M$ that according to \cite{Spruit2015}'s suggestion is the maximum one. Also we see that the hypothesis made in \eq{eq_discont_T} maximizes the value of $\Delta M/M$. 

When the convective core grows, a parameter-free alternative approach is to use the entrainment velocity, obtained from numerical simulations, given by \cite{Meakin2007a}. It provides the growth rate of the core boundary  $\nu$. It is given by $\nu = A Ri_B^{-n}$ with $\ln A=0.027\pm0.38$ and $n=1.05\pm0.21$.  We will define the Richardson number as in Cristini et al (2017) by 
\begin{align}
Ri_B = \frac{l \int_{0}^{r_D+\Delta r} N^2 {\rm d}r}{V_M^2} \, , 
\end{align}
where $V_M$ is the mean velocity at the location of the discontinuity, $\Delta r$ is the penetration depth in the stable region, and $l$ is a length scale for the turbulent motions, which is often taken to be the horizontal integral scale of the turbulence near the interface, according \cite{Meakin2007a}, so that $l \approx r_D \sin \theta_M (r_D)$. The lower boundary of the integral is not very important as the main contribution will come from the sub-adiabatic layers below the chemical discontinuity. 
When computing the integral, we should not forget that $N^2$ is a delta function at the location of the chemical composition discontinuity and then varies slowly in the stable region where  $N^2>0$. As the penetration depth will be very small compared to the density scale height, we have 
\begin{align}
\int_{r_D}^{r_D+\Delta r} N^2 {\rm d}r \approx g(r_D) \ln \frac{\rho_-}{\rho_+} \, , 
\end{align}
where $g$ is the gravitational acceleration and  $\rho_-$ and $\rho_+$ are respectively the density below and above the chemical composition discontinuity so that 
\begin{align}
Ri_B = \frac{l \left[\int_{0}^{r_D} N^2 {\rm d}r + g(r_D) \ln \frac{\rho_-}{\rho_+}\right]}{V_M^2} \, .
\end{align}
If when the discontinuity coincides with the point where $L=L_R$ (as assumed in stellar evolution without overshooting) the entrainment rate is larger than required by stellar evolution ($Ri_B$ too small), the sub-adiabatic layer with $\nabla < \nabla_a$  ($N^2>0$) will grow until $\int_0^{r_D} N^2 {\rm d}r$ becomes sufficiently positive and $V_M$ sufficiently small to  have the required value of the entrainment rate (this will correspond to a Richardson number only slightly smaller than $1/4$). If without overshooting the entrainment rate is too small ($Ri_B$ too large), then we have an evolution with undershooting and the discontinuity should be moved inwards in order to reduce the size of the sub-adiabatic region.

\section{Conclusion}
\label{section_9}

Up to now, there have been mainly  three ways of handling the overshooting region at the upper boundary of convective zones. First, in most of the current existing codes, its extent is considered as a free parameter (generally a fraction of the local pressure scale height) and more importantly the temperature gradient is arbitrarily taken as either equal to the adiabatic one or to the radiative one. Second, higher-order methods based on Reynolds stress models naturally provide an estimate of the overshooting. However, such models are subject to many uncertainties and are quite difficult to implement in stellar evolutionary codes. Third, based on numerical simulations \citep[see][]{Meakin2007a,Arnett2009,Arnett2011,Viallet2013}, \cite{Arnett2015} (Sect. 3.8)  propose an alternative model to account for the braking at the upper edge of convective cores  and more generally to compute the structure of convective zones. They named it the 321D algorithm. 

In this work, we have developed a theory of convection based on a plume model, which takes also into account the counter flow. We have also introduced a new equation that closes the system. The theory is comparable with a non-local mixing length theory but with a somewhat different interpretation of the mixing length. Its main advantage is that it allows the computation of the departure of the temperature gradient from the adiabatic value in the whole convective core, that is to say also in the overshooting layers. 
This finally provides eight differential equations (including the usual ones: the mass conservation, the hydrostatic equilibrium, and the thermal energy conservation) and essentially two algebraic relations for ten unknowns. It is worth noticing that the set of equations is complete if we assume that the number of regions with down- and up-moving flows is equal. If not, the number of down-moving regions is a free parameter.  Interestingly enough, we do not have to make several hypotheses done in the classical plume theory, such as Taylor's advection (or entrainment) hypothesis, which is replaced by a closure based on mean-field equations.  Of course our theory still contains a few free parameters, some of which are the same as in the LMLT while others are related to the assumed  geometry. In this respect, 3D hydrodynamical simulations are certainly valuable tools to help constrain those parameters. We also  notice that the 321D approach as proposed by \cite{Arnett2015} is certainly more general since it does not require flow assumptions. Comparisons between our approach and the 321D model are thus desirable in the near future.

The equation giving the departure from the adiabatic gradient shows that at the bottom of the overshooting zone the temperature gradient will be close to the adiabatic value, but that when the distance to the bottom becomes a significant fraction of the pressure scale height the temperature gradient becomes strongly sub-adiabatic and progressively reaches the radiative value. It is what is also found by numerical simulations \citep{Brummell2002}, with Xiong’s code \citep[see][and the reference therein]{Xiong2001,Zhang2012}, or by use of semi-analytical models \citep{Rempel2004}. 
Also the development of the theory allowed us to show rigorously, that is without any approximation, that it is necessary that the convective kinetic energy flux is not null everywhere in order to have overshooting at any boundary of a convective zone. This implies that all theories developed so far that neglect that term get results because they neglect at least one of the basic equations of the problem, very often the convective kinetic energy conservation.  Consequently, their results must be considered with caution. 
Indeed, as discussed in Appendix B and shown in \cite{Arnett2015}, the braking region must always be sub-adiabatic in order to have enough braking and this region does not exist in the LMLT. 
 
Finally, we emphasize that our theory is not valid in the surface layers of stars where convective cells form (i.e. at the photosphere). However, it could be adapted to convective envelopes provided numerical simulations can provide the boundary conditions below the strongly non-adiabatic layers. 
We also point out that the bulk of our work focused on the case of shrinking convective cores. For an expanding core, the problem is much more difficult because one has to determine the mixing rate at the location of the chemical composition discontinuity. We did not solve that problem but we introduced an extra parameter the minimum value of which can be obtained from an evolution without overshooting. This will at least provide the order of magnitude of that parameter. This is certainly an issue that deserves more theoretical development. However, an alternative approach based on results of numerical simulations and which is parameter-free has been given by \cite{Meakin2007a}.

\begin{acknowledgements}
We thank Arlette Noels for providing us with the models required to compute Table~\ref{table:1}. 
\end{acknowledgements} 

\newpage

\bibliographystyle{aa}


\appendix 

\section{Computation of the coefficients $J_{i,M}, J_{i,D}$ , and $J_{i,D}^\prime$}
\label{appendixA}

In this appendix, we provide the coefficients $J_{i,M}, J_{i,D}$ ,and $J_{i,D}^\prime$ that appear from angular integration on a spherical surface or over a plume horizontal section. The computation of those coefficients is simple even if somewhat lengthy. It requires only the use of simple fundamental trigonometric functions. 

\subsection{The coefficients $J_{i,M}, J_{i,D}$}

In the first hypothesis discussed  at the end of Sect.~\ref{defs_horizontal}, we have $J_{i,M}=J_{i,D}$. We thus start by defining the parameter $\zeta = \pi / (2 \theta_M)$, so that one obtains 

\begin{itemize}
        \item for $J_{0,M}$
        \begin{align}
                J_{0,M}  &= \int_0^{\theta_M} \cos \left(\zeta \theta \right) \sin \theta \, {\rm d}\theta = \frac{1-\zeta \sin \theta_M}{1-\zeta^2} \, , 
        \end{align}
        and, in the limit of small angle $\theta_M$, one has 
        \begin{align}
                \label{small_JOM}
                J_{0,M}  &\simeq \zeta^{-2} \left( \frac{\pi}{2} - 1\right) \, .
        \end{align}
        
        \item for $J_{1,M}$
        \begin{align}
                J_{1,M}  &= \int_0^{\theta_M} \cos^2 \left(\zeta \theta \right) \sin \theta \, {\rm d}\theta \nonumber \\
                &= \frac{1}{2} \left[ (1-\cos \theta_M) + \frac{(1+\cos \theta_M)}{1-(2\zeta)^2}\right]
        \end{align}
        and, in the limit of small angle $\theta_M$, one has 
        \begin{align}
                J_{1,M}  \simeq \left(\frac{1}{2 \zeta}\right)^2 \left[\frac{\pi^2}{4}-1\right] \, .
        \end{align}
        
        \item for $J_{2,M}$
        \begin{align}
                &J_{2,M}  = \int_0^{\theta_M} \left(\zeta \theta \right) \sin \left(\zeta \theta \right) \sin \theta \, {\rm d}\theta 
                = \frac{\zeta / 2}{\left[1-\zeta^2\right]^{2}} \nonumber \\
                &\times \left\{ -4 \zeta + 2 \sin \left(\theta_M\right) \left(1+\zeta^2\right)
                - \frac{\pi}{\zeta} \cos\left(\theta_M\right)  \left(1-\zeta^2\right)
                \right\}
        \end{align}
        and, in the limit of small  angle $\theta_M$, one has 
        \begin{align}
                J_{2,M}  \simeq \zeta^{-2} \left(\pi- 2\right) \, .
        \end{align}
        
        \item for $J_{3,M}$
        \begin{align}
                J_{3,M} &= \int_0^{\theta_M} \left(\zeta \theta \right)^2 \cos \left(\zeta \theta \right) \sin \theta \, {\rm d}\theta  \\ 
                &= \frac{\zeta^2}{\left(\zeta-1\right)^3} \left[\cos \theta_M \, \mathcal{F}_1 + \sin \theta_M \, \mathcal{F}_2 
                + \frac{\pi^2}{2\zeta} \left(3\zeta^2+1\right) \right]\nonumber
        \end{align}
        with
        \begin{align}
                \mathcal{F}_1 &= 2\pi \left(1-\zeta^2\right) \\
                \mathcal{F}_2 &= \left[\zeta^3+3\zeta\right] \left[\left(\frac{\pi}{2}\right)^2 + \left(\frac{\pi}{2\zeta}\right)^2 -2\right] - 
                2\zeta \left[3\zeta^2+1\right]
        \end{align}
        and, in the limit of small angle $\theta_M$, one has 
        \begin{align}
                J_{3,M}  \simeq \zeta^{-2} \left[\left(\frac{\pi}{2}\right)^3 - 6 \left(\frac{\pi}{2}-1\right)\right] \, .
        \end{align}
        
        \item for $J_{4,M}$
        \begin{align}
                J_{4,M} &= \int_0^{\theta_M} \cos^3 \left(\zeta \theta \right) \sin \theta \, {\rm d}\theta \nonumber \\ 
                &= \frac{1}{4} \left\{ 3 \frac{1-\zeta \sin \left(\theta_M\right)}{1-\zeta^2} 
                +\frac{1+3\zeta \sin \left(\theta_M\right)}{1-9\zeta^2}  \right\}
        \end{align}
        and, in the limit of small angle $\theta_M$, one has 
        \begin{align}
                J_{4,M}  \simeq \zeta^{-2} \left[\frac{\pi}{3} - \frac{7}{9}\right] \, .
        \end{align}
        
        \item for $J_{5,M}$
        \begin{align}
                \frac{4}{\zeta} J_{5,M} &= \frac{4}{\zeta} \int_0^{\theta_M} \left( \theta \zeta\right) \cos^2 \left(\zeta \theta \right) \sin \left(\zeta \theta \right) \sin \theta \, {\rm d}\theta  \\ 
                &= \frac{\left(\zeta^2+1\right) \sin\left(\theta_M\right)-2\zeta + \frac{\pi}{2\zeta} \left(\zeta^2-1\right) \cos\left(\theta_M\right) }{\left(\zeta^2-1\right)^2} \nonumber \\
                &- \frac{\left(9\zeta^2+1\right) \sin\left(\theta_M\right)-6\zeta - \frac{\pi}{2\zeta}\left(9\zeta^2-1\right)\cos\left(\theta_M\right) }{\left(9\zeta^2-1\right)^2}\nonumber
        \end{align}
        and, in the limit of small angle $\theta_M$, one has 
        \begin{align}
                3 J_{5,M}  \simeq 2 \zeta^{-2} \left[\frac{\pi}{3} - \frac{7}{9}\right]
        \end{align}
        
        \item for $J_{6,M}$
        \begin{align}
                & J_{6,M} = 2\int_0^{\theta_M} \left( \theta \zeta\right) \cos \left(\zeta \theta \right) \sin \left(\zeta \theta \right) \sin \theta \, {\rm d}\theta  \\ 
                &= \frac{\pi\zeta}{\left(4\zeta^2-1\right)^2} \left[-\frac{4\zeta}{\pi} \left(\cos \left(\theta_M\right)+1\right) 
                +\sin \left(\theta_M\right) \left(4\zeta^2 -1\right)\right] \nonumber
        \end{align}
        and, in the limit of small angle $\theta_M$, one has 
        \begin{align}
                J_{6,M}  \simeq \frac{1}{8 \zeta^2}  \left[\pi^2-4\right] \, .
        \end{align}
\end{itemize}
 We also have 
\begin{align}
        S_M = \frac{N}{2} \left(1-\cos \theta_M\right) \, .
\end{align}

When we assume that the geometry of the descending regions is the same as that of the plume, the coefficients $J_{i,D}$ are given by the same formulae as the $J_{i,M}$ (and $\theta_M$ becomes $\theta_D$). However, when we assume that each plume is surrounded by a hollow cone with an opening between $\theta_M$ and $\theta_{M,D}$ where the gas is flowing downwards, the $J_{i,D}$ are given by the different formulae that we write $J_{i,D}^\prime$. 

\subsection{The coefficients $J_{i,D}^\prime$}

In the following, we will consider only the coefficients $J_{i,D}^\prime$ that are required in the integration over a spherical surface. 

First, we have for $S_D$
\begin{align}
        S_D = \int_{\theta_M}^{\theta_{M,D}} \; \sin \theta \, {\rm d}\theta = \frac{N}{2} \left(\cos\theta_M - \cos\theta_{M,D}\right)
\end{align}
and the condition $S_M+S_D=1$ leads to
\begin{align}
\label{eq_A19}
        \frac{N}{2} \left(1-\cos \theta_{M,D}\right) = 1 \, .
\end{align}

The condition $\overline{\rho \vec V} = 0$ implies (see Eq.~\ref{mass_flux_AD2}) 
\begin{align}
        \label{tmp_appendix}
        &J_{0,M} + A_D J_{0,D}^\prime = \\
        &\int_{0}^{\theta_M} \; \cos\left(\frac{\pi}{2}\frac{\theta}{\theta_M}\right) \sin\theta {\rm d}\theta + A_D \int_{\theta_M}^{\theta_{M,D}} \;
        \sin \left(\pi \frac{\theta-\theta_M}{\theta_{M,D}-\theta_M}\right) \sin\theta {\rm d}\theta =0 \nonumber
\end{align}
with
\begin{align}
        J_{0,D}^\prime = \frac{\pi \left(\theta_{M,D}-\theta_M\right)}{\pi^2-\left(\theta_{M,D}-\theta_M\right)^2} \; \left[\sin(\theta_{M,D})+\sin(\theta_M)\right] \, .
\end{align}
 Therefore, \eq{tmp_appendix} gives $A_D$. If $\theta_{M,D}$ and $\theta_M$ are small, one has
\begin{align}
        J_{O,D}^\prime \simeq \frac{\left(\theta_{M,D}^2-\theta_M^2\right)}{\pi} \, .
\end{align}
 Consequently, using \eq{small_JOM}, one gets
\begin{align}
        A_D \simeq \frac{4}{\pi} \left(1-\frac{\pi}{2}\right) \left(\frac{\theta_M}{\theta_{M,D}-\theta_M}\right)^2 
        = \frac{4}{\pi} \left(1-\frac{\pi}{2}\right)  \frac{S_M}{S_D} \, .
\end{align}

Now we consider the angular integrals $G_i$ in \eq{defs_Gi}. This gives
\begin{itemize}
        \item for $G_1$
        \begin{align}
                G_1 = \frac{N}{2} J_{1,M} + \frac{N}{2} A_D^2 J_{1,D}^\prime
        \end{align}
        where 
        \begin{align}
                J_{1,D}^\prime &= \int_{\theta_M}^{\theta_{M,D}} \, \sin^2 \left(\pi \frac{\theta-\theta_M}{\theta_{M,D}-\theta_M}\right) 
                \sin \theta \, {\rm d}\theta \nonumber \\
                &=\frac{2\pi^2}{4\pi-\left(\theta_{M,D}-\theta_M\right)^2} \left(\cos \theta_M - \cos \theta_{M,D}\right) \, 
        \end{align}
        which, for small values of $\theta_{M,D}$ and $\theta_M$, reduces to
        \begin{align}
                J_{1,D}^\prime \simeq \frac{\pi^2 \left(\theta_{M,D}^2-\theta_M^2\right)}{4\pi^2 - \left(\theta_{M,D}^2-\theta_M^2\right)}
                \simeq \frac{\left(\theta_{M,D}^2-\theta_M^2\right)}{4} \, . 
        \end{align}
        
        \item for $G_3$ and $G_4$
        \begin{align}
                G_3 &= \frac{N}{2} J_{4,M} + \frac{N}{2} A_D^3 J_{4,D}^\prime \\
                G_4 &= \frac{N}{2} J_{4,M} + \frac{N}{2} \left\vert A_D^3 \right\vert J_{4,D}^\prime
        \end{align}
        where 
        \begin{align}
                J_{4,D}^\prime &= \int_{\theta_M}^{\theta_{M,D}} \, \sin^3 \left(\pi \frac{\theta-\theta_M}{\theta_{M,D}-\theta_M}\right) 
                \sin \theta \, {\rm d}\theta \nonumber \\
                &= \frac{6\pi^2 \left(\theta_{M,D}-\theta_M\right)\left(\sin\theta_{M,D}+\sin\theta_M\right)}{\left[\pi^2-\left(\theta_{M,D}-\theta_M\right)^2\right] \left[9\pi^2-\left(\theta_{M,D}-\theta_M\right)^2\right]} \, , 
        \end{align}
        which, for small values of $\theta_{M,D}$ and $\theta_M$, reduces to
        \begin{align}
                J_{4,D}^\prime \simeq \frac{2}{3\pi} \left[ \theta_{M,D}^2-\theta_M^2 \right] \, . 
        \end{align}
\end{itemize}

\section{Influence of the sign of $L_K$ on the extent of the overshooting}
\label{appendixB}

From \eq{mechanical_sphere_final} and \eq{luminosity_sphere_final} 
\begin{align}
        \frac{1}{r^2} \deriv{}{r} \left(\frac{k}{2} r^2 \left(\rho V_r^3\right)_0 G_3 \right) &= 
        \frac{Qg}{C_p \overline{T}} \left[\frac{\overline{L}-\overline{L}_R}{4\pi r^2}-\frac{k}{2} \left(\rho V_r^3\right)_0 G_3\right] \nonumber \\
        &-\frac{k\left(\rho V_r^3\right)_0}{l} G_4 \, .
\end{align}
The latter equation can be recast such that 
\begin{align}
        \label{tmp_eq_appendix2}
        \deriv{y}{r} + \left[\frac{2}{l}\frac{G_4}{G_3}+\frac{\nabla_a}{H_p}\right] y =
        \frac{\nabla_a}{H_p} \frac{L-L_R}{4\pi} 
\end{align}
where 
\begin{align}
        y = \frac{k}{2} r^2 \left(\rho V_r^3\right)_0 G_3.
\end{align}
When $G_3>0$ ($F_{K,r}>0$), then $y>0$ and \eq{tmp_eq_appendix2} shows that ${\rm d}y/{\rm d}r<0$ when $L<L_R$, that is to say above the point where $L=L_R$, as $G_4>0$. This implies that braking starts when the convective luminosity is negative, that is to say when the temperature fluctuation of rising material is negative, which is only possible if the temperature gradient is sub-adiabatic. This is also what is found in numerical simulations \citep[see][]{Arnett2015}. This seems to exclude the second possibility discussed next. Then we are sure to obtain a solution $y=0$.
On the other hand, if $G_3<0$ ($F_{K,r}<0$) then $y<0$ and it is necessary that ${\rm d}y/{\rm d}r>0$  above the point where $L=L_R$ to have a point where $y=0$. However, the only term which contributes to make the derivative positive is $\nabla_a y /H_p$  and as long as  $(L-L_R)/4\pi-y<0$, the derivative is negative, $y$  decreases as the radius grows, and it is impossible to have $y=0$. Consequently, when $G_3<0$ the existence of a solution seems problematic or, if it exists, it will be found for a very small overshooting.

\end{document}